\documentclass[preprint,amsmath,amssymb,aps,a4]{revtex4-2}
\usepackage{graphicx}
\usepackage{bm}
\usepackage{txfonts}
\usepackage{hyperref}
\usepackage[english]{babel}

\usepackage[autostyle, english = american]{csquotes}
\MakeOuterQuote{"}
\date{\today}
\newcommand{\be}{\begin{eqnarray}}
\newcommand{\ee}{\end{eqnarray}}

\newcommand{\bfk}{{\bf k}_{\perp}}
\newcommand{\bfkpr}{{\bf k}_{\perp}^\prime}
\newcommand{\bfki}{{\bf k}_{\perp i}}
\newcommand{\bfb}{{\bf b}_{\perp}}

\newcommand{\bfP}{{\bf P}_{\perp}}

\newcommand{\bfpi}{{\bf p}_{\perp i}}
\newcommand{\bfpip}{{\bf p}_{\perp i}^\prime}

\newcommand{\Dp}{{\bf \Delta}_{\perp}}

\newcommand{\lipr}{{\lambda_{i}}^\prime}

\newcommand{\bfr}{{\bf r}_{\perp}}
\usepackage{xcolor}
\begin{document}
\title{Generalized parton distributions for the lowest-lying octet baryons}
\author{Navpreet Kaur}
\email{knavpreet.hep@gmail.com}
\affiliation{Department of Physics, Dr. B.R. Ambedkar National
Institute of Technology, Jalandhar, 144008, India}

\author{Harleen Dahiya}
\email{dahiyah@nitj.ac.in}
\affiliation{Department of Physics, Dr. B.R. Ambedkar National
Institute of Technology, Jalandhar, 144008, India}

\date{\today}%
\begin{abstract}
We have presented a pathway to map the internal structure of lowest-lying octet baryons by using one-loop quantum fluctuations of a fermion state in Yukawa theory. The structural interpretation of baryons carrying different amount of strange content has been exemplified by employing a quark-scalar diquark model for studying the Generalized Parton Distributions (GPDs) among the members of octet baryons. Different possible combinations of quark-scalar diquark pairs have been considered. The quark helicity independent chiral even GPDs in momentum as well as impact parameter space have been studied  for purely transverse momentum transfer (zero skewness). We have also extended our calculations to explore and visualize the distinct behavior of charge distributions in coordinate space as well as charge densities in impact parameter space. Furthermore, magnetization density in impact parameter space has also been investigated. Quantitative as well as qualitative analysis of all the distributions has been carried out.
\end{abstract}
%
\maketitle
%
%
\section{Introduction\label{secintro}}
The layers of a material have been unfolded from an atom to the present day picture of partons (quarks and gluons). Physicists have been trying to understand the scenario of partons inside the complex composite system of hadrons from the past few decades. However, quantum physics closes the voids in our knowledge of physics to give us a more detailed picture of such systems. At present, we are familiar with the Lagrangian of Quantum chromodynamics (QCD) acting as a baseline to examine the complex hadron structure in terms of the quark and gluon fields stated by its Fock components. To have a deeper understanding, these quark and gluon fields can be tossed on the states of a hadron to obtain the matrix elements which are interpreted as the wave functions of a hadron \cite{Brodsky:1989pv}. Since a direct extraction of the hadron wave function is not possible experimentally, some phenomenological functions are accessed to describe the complex hadronic internal structure. For a long time, non-perturbative functions such as form factors \cite{Sterman:1997sx, JeffersonLabHallA:2001qqe, SAMPLE:2003wwa, Punjabi:2005wq}, parton densities \cite{Leader:2001bp} and distribution amplitudes \cite{Lepage:1980fj} have been used to study the internal structure of hadrons. The matrix elements of electromagnetic and weak currents are interpreted as form factors rich in explaining the transverse structure of the hadrons \cite{Gayou:2001qt}. They are further related to the electric charge and the anomalous magnetic moment of the composite system. The forward matrix elements of the partonic field separated by light like distances interpret the momentum space and are termed as usual parton densities. They explain the longitudinal structure of the hadrons whereas, the matrix elements of the light-cone operators give the distribution amplitudes. These non-perturbative functions can be accessed by elastic as well as inelastic scattering processes \cite{Lepage:1980fj,Altarelli:1972sw}.

The information about the partonic structure of the hadrons is encoded in the Generalized Parton Correlation Function (GPCF) which gives the Generalized Transverse Momentum Distributions (GTMDs) after integrating over one light-cone component of the quark momentum \cite{Meissner:2009zz}. Setting the total momentum transferred as zero, one can get the Transverse Momentum Distributions (TMDs) \cite{Barone:2001sp,Mulders:1995dh} from GTMDs and depict a three dimensional (3D) representation of a composite system in terms of longitudinal momentum fraction and the transverse momentum carried by the quark or gluon. On integrating over the partonic transverse momentum, one can get a generalized function: Generalized Parton Distribution (GPD) \cite{Muller:1994ses,Ji:1996ek,Radyushkin:1996nd,Goeke:2001tz} in which all above mentioned non-perturbative functions get fused. GPD also has a capability to portrait a 3D picture of a composite system as it describes the distribution of quarks or gluons inside the hadron \cite{Ji:2004gf}. It is a function of longitudinal momentum fraction carried by the quark or gluon $x$, longitudinal momentum transferred in the process $\zeta$ and square of the total momentum transferred $t$ \cite{Diehl:2003ny}. This can also be extended to the transverse space. The merits of contribution in the development of GPD goes to Leipzig \cite{Geyer:1985vw,Braunschweig:1985nr,Dittes:1988xz}, Bartels and Loewe \cite{Bartels:1981jh}, Collins \cite{Collins:1996fb}, Ji \cite{Ji:1996ek} and Radyushkin \cite{Radyushkin:1996nd,Radyushkin:1996ru} who put forward the description of the mathematical tool of GPDs. The importance of the 3D projection of GPD was further recognized after the representation of GPD in impact parameter space \cite{Burkardt:2000za}.

GPDs are the non-local and non-forward matrix elements, accessible in exclusive processes like Deeply Virtual Compton Scattering (DVCS) \cite{Collins:1998be, BessidskaiaBylund:2022qgg, Braun:2022qly} and Meson Production (DVMP) \cite{CLAS:2022iqy, Goloskokov:2007nt, Vanderhaeghen:1999xj, Goloskokov:2005sd}. $H_{X}^{q}(x,0,\Dp)$ is an unpolarized GPD describing the distribution of a quark $q$ or gluon $g$ inside an unpolarized hadron target $X$, whereas $E_{X}^{q}(x,0,\Dp)$ is a transversely polarized GPD which gives a measure of the transverse deformation in an active quark $q$ or gluon $g$ distribution inside a unpolarized hadron target $X$. Fourier transformation of GPDs with respect to the transverse momentum transfer leads us to the space of impact parameter or to the transverse plane. These distributions are termed as “Impact Parameter Dependent Parton Distribution Functions”, IPDPDFs \cite{Boffi:2007yc}. Impact parameter distance assesses the transverse distance of an active quark from the center of momentum of the composite system of a hadron. Charge density is linked to Dirac form factors and their connection can be determined from the current density \cite{Miller:2010nz}. As the form factors can be obtained from GPDs \cite{Belitsky:2005qn}, one can make use of $H_{X}^{q}(x,0,\Dp)$ to study the charge densities in impact parameter space which is Fourier conjugate to the momentum space \cite{Selyugin:2009ic}. Along with the study of charge density in impact parameter space, one can also examine the charge distributions in coordinate space  \cite{Kumar:2014coa} to analyze the distinct behavior of these distributions in two different spaces. Another application of form factors is the magnetization density which is linked to Pauli form factor and it can be obtained from GPD $E_{X}^{q}(x,0,\Dp)$ \cite{Miller:2010nz}.

About a half century ago, the existence of quarks inside the nucleons was announced experimentally. Till now, a lot of work has been done theoretically \cite{Sharma:2023wha, sstwist4, Djukanovic:2023beb, Liang:2014hca} as well as experimentally \cite{SND:2023fos, Riedl:2022bwc, HERMES:2001bob} to understand the behavior of quarks inside the nucleons, the genesis of mass \cite{GlueX:2019mkq,Kharzeev:2021qkd} and spin of nucleons \cite{Riedl:2022pad} which are the members of a baryon octet. The data of DVCS process from CLAS and Hall A collaboration at Jefferson lab (JLab) has been examined within GPD formalism to study the dependency of charge radius of a proton on $x$ and  the constraints on Compton form factors (CFFs) \cite{Dupre:2017hfs} have been figured out. More precise study of GPDs in the 12 GeV program at Jefferson lab will help gather an extensive data to scrutinize the internal structure of a proton \cite{Achenbach:2023pba}. The results released from HERA and HERMES at DESY will also provide substantial information about the nucleon tomography \cite{Moutarde:2019tqa}. Switching to theoretical studies, GPDs of proton have been studied using lattice QCD \cite{Lin:2021brq, Bhattacharya:2023nmv}, basis light-front quantization \cite{Kaur:2023lun}, universal moment parameterization \cite{Guo:2023ahv}, light-cone diquark model \cite{Lu:2010dt}, light-front constituent quark model \cite{Lorce:2011dv} and Nambu-Jona-Lasinio (NJL) model \cite{Freese:2020mcx}. 

In light of the above developments, it becomes essential to step ahead and extend the studies for other baryons having same spin-parity quantum number $J^{P}=(\frac{1}{2})^{+}$. The lowest-lying baryon octet includes families of $N$, $\Sigma$, $\Xi$ and a $\Lambda$ which need to be scrutinized. TMDs of strange and charm quarks \cite{Zhu:2023nhl}, electroweak properties \cite{Zhang:2016qqg} and quark flavor decomposition \cite{Dahiya:2015wqa} for the octet baryons have been studied recently. Other than this, no theoretical and experimental data about baryon octet is currently available in the context of bilinear current vectors. Experimental study of DVCS processes to scrutinize the strangeness containing baryons ($\Sigma$, $\Xi$ and $\Lambda$) is not available yet as the life time of strange baryons is very small. Study of strange baryons is of paramount importance as their behavior is interlinked to the properties of neutron stars and strange stars \cite{Wang:2005vg,Vidana:2018bdi}. The behavior analysis of strange particles can also be fruitful to examine hypernuclear physics \cite{Feliciello:2015dua}. Therefore, our small step to probe the  strange baryons in octet by making the use of quark-scalar diquark model with the tool of GPD will act as a  preliminary investigation for a more extensive understanding.

The present paper is organized as follows. In Section \ref{secmodel}, kinematics of the DVCS process and detailed model description used in this work has been presented. Numerical parameters employed in the calculation have been listed in Section  \ref{secnumpar}. Section \ref{secgpd} is devoted to the elucidation of GPD in terms of bilinear currents to obtain its explicit form. In Section \ref{secipdpdf}, parton distributions in impact parameter space have been deduced and presented. Charge distribution in coordinate space and charge density in impact parameter space have been evaluated in Section \ref{secchargedis}. Contrasting nature of charge distribution and density have also been shown in the same section. Further, in Section \ref{secmagden}, magnetization density of baryons containing different strange content has been discussed. We summarize and conclude our findings in Section \ref{seccon}.

	\section{Kinematics and Model Description \label{secmodel}}
    DVCS in the limit of large initial photon virtuality is a crucial scattering process to determine the GPDs. For comprehensiveness, we discuss the required kinematics of the following DVCS process \cite{Brodsky:2000rt, Brodsky:2000xy}
	\begin{equation}
		\gamma^{\ast}(q) + p(P) \rightarrow \gamma(q^{\prime})+p(P^{\prime}),
	\end{equation}
	where $P$ and $P^{\prime}$ are the initial and final momentum of a hadron state respectively. We choose a frame whose light-cone coordinates are
	\begin{eqnarray}
	P &=& \bigg(P^{+},{\bf 0_\perp},\frac{M^2}{P^{+}} \bigg) \, , \\
	P^\prime &=& \bigg((1-\zeta)P^{+}, -\Dp, \frac{{M^2}+\Dp^2}{(1-\zeta)P^{+}} \bigg) \, ,
	\end{eqnarray}
	where M and $\zeta$ denote the mass of a hadron and skewness respectively. $\Dp$ represents the four momentum transfer from a hadron which is given by
	\begin{equation}
		\Delta = P-P^{\prime} = \bigg(\zeta P^{+}, \Dp, \frac{t+\Dp^2}{\zeta P^+} \bigg),
	\end{equation}
	with 
	\begin{equation}
		t=2P\cdot\Delta=-\frac{\zeta^{2} M^{2}+\Dp^2}{1-\zeta}.
	\end{equation}
	For zero skewness, we have $t=\Delta^{2}=-\Dp^{2}$.
	
We have adopted a QCD inspired quark-scalar diquark model of a hadron which represents a simplistic panorama of an active quark and a spectator of a diquark $n$. Following Ref. \cite{Brodsky:2002cx}, we consider a hadron of mass $M$ consisting of a spin-$\frac{1}{2}$ active quark and a spin-$0$ spectator with respective masses $m$ and $\lambda$. Based on the one-loop quantum fluctuations of the Yukawa theory, the Yukawa Lagrangian density is expressed as \cite{Brodsky:2000ii} 
	\begin{eqnarray}
	\mathcal{L}=\frac{i}{2}
	[\bar{\psi}\gamma^{\mu}(\partial_{\mu}\psi)-(\partial_{\mu} \bar{\psi})\gamma^{\mu}\psi)]-m\bar{\psi}\psi+\frac{1}{2} (\partial^{\mu}\phi)(\partial_{\mu}\phi)-\frac{1}{2}\lambda^{2}\phi\phi+g\phi\bar{\psi}\psi,
	\end{eqnarray}	
and the Noether current defining the corresponding conserved energy-momentum tensor in terms of a general coordinate tranformation is given as
	\begin{eqnarray}
	\mathcal{T}^{\mu \nu}=\frac{i}{2} [\bar{\psi}\gamma^{\mu}\overrightarrow{\partial}^{\nu}\psi-\bar{\psi}\gamma^{\mu} \overleftarrow{\partial}^{\nu}\psi]+(\partial^{\mu}\phi)(\partial^{\nu}\phi)- g^{\mu \nu} \mathcal{L} \, .
	\end{eqnarray}
	A hadron is an eigenstate of invariant light-cone Hamiltonian,  $H_{LC}^{QCD}=P^{+} P^{-}-{\bf P_\perp^2}$ with an invariant mass square as its eigenvalue. Here, the momentum generators, $P^{+}$ and $\bfP$, are kinematical quantities as they are free from an interaction. On the other hand, the generator $P^{-}$, governs the light-cone time translation. In the light-cone dynamics, instead of an ordinary time, the fields are quantized at fixed light-cone time, $\tau=t+z/c$ \cite{Dirac:1949cp} and the expansion of a hadron state can be jotted down on the complete basis of free $|\mathcal{N}\rangle$ Fock states of constituents as follows
	\begin{eqnarray}
		|\psi_{A}(P^{+},{\bf P_\perp^2)}\rangle &=& \sum_{\mathcal{N}} \prod_{i=1}^{\mathcal{N}} \frac{dx~  d^2\bfki}{2(2\pi)^3\sqrt{x_{i}}} \, 16 \pi^{3} \, \delta \bigg(1-\sum_{i=1}^{\mathcal{N}} x_{i}\bigg) \, \delta^{(2)} \bigg(\sum_{i=1}^{\mathcal{N}}\bfki\bigg) \nonumber \\		
		&\times& \psi_{\mathcal{N}}(x_{i},\bfki,\lambda_{i})|\mathcal{N}; x_{i} P^{+},x_{i}\bfP+\bfki,\lambda_{i}\rangle \, ,
	\end{eqnarray}
	where $\bfki$, $\lambda_{i}$ and  $x_{i}=k_{i}^{+}/P^{+}$ symbolize the light-cone intrinsic transverse momentum, helicity and longitudinal momentum fraction acquired by an \textit{i}th component of a hadron, respectively. We also avail the light-cone gauge, $A^{+}=0$. The frame independent light-cone wave function $\psi_{\mathcal{N}}^{X}$ represents the projection of a hadron state on the Fock state $|\mathcal{N}\rangle$. The wave function of each Fock-state of a hadron with total spin $J^{z}=\pm\frac{1}{2}$ and is denoted by $\psi_{\mathcal{N}}^{J^{z} X}(x_{i},\bfki,\lambda_{i})$ with
	\begin{equation}
		k_{i}=(k_{i}^{+},k_{i}^{-},\bfki)=\bigg(x_{i}P^{+},\frac{\bfki^{2}+m_{i}^{2}}{x_{i}P^{+}},\bfki\bigg) \, .
	\end{equation}
	The states of $\mathcal{N}$-particles are normalized as 
	\begin{equation}
		\langle \mathcal{N}; p_{i}^{\prime +}, \bfpip, \lambda_{i}^{\prime} |\mathcal{N};  p_{i}^{+},\bfpi,\lambda_{i}\rangle = \prod_{i=1}^{\mathcal{N}} 16 \pi^{3} \, p_{i}^{+} \, \delta(p_{i}^{\prime +}-p_{i}^{+}) \, \delta^{(2)} (\bfpip-\bfpi) \, {\delta}_{\lipr \lambda_{i}} \, .
	\end{equation}
	For a hadron with $J^{z}=+\frac{1}{2}$, the two particle Fock state has two possible spin combinations with
	\begin{eqnarray}
	|\psi_{2 \rm{particle}}^{\uparrow X}(P^{+},\bfP=\bf{0}_{\perp})\rangle & =& \int \frac{dx~  d^2\bfki}{2(2\pi)^3\sqrt{x(1-x)}} \, \bigg[\psi_{+\frac{1}{2}}^{\uparrow X}(x,\bfk)\bigg|+\frac{1}{2};x P^{+},\bfk\bigg\rangle \nonumber \\
	&+& \bigg[\psi_{-\frac{1}{2}}^{\uparrow X}(x,\bfk)\bigg|-\frac{1}{2};x P^{+},\bfk\bigg\rangle \bigg] \, ,
	\end{eqnarray}
	where
	\begin{eqnarray}
	\psi_{+\frac{1}{2}}^{\uparrow X}(x,\bfk) &=& \bigg(M_{X}+\frac{m_{q}}{x}\bigg) \, \varphi_{X} \, , \nonumber \\
	\psi_{-\frac{1}{2}}^{\uparrow X}(x,\bfk) &=& -\frac{(k^{1}+i k^{2})}{x} \, \varphi_{X}\, .
	\label{WFSPH}
	\end{eqnarray}
	Similarly, for $J^{z}=-\frac{1}{2}$ hadron, the two particle Fock state is
	\begin{eqnarray}
	|\psi_{2 \rm{particle}}^{\downarrow X}(P^{+},\bfP=\bf{0}_{\perp})\rangle & =& \int \frac{dx~  d^2\bfki}{2(2\pi)^3\sqrt{x(1-x)}} \, \bigg[\psi_{+\frac{1}{2}}^{\downarrow X}(x,\bfk)\bigg|+\frac{1}{2};x P^{+},\bfk\bigg\rangle \nonumber \\
	&+& \bigg[\psi_{-\frac{1}{2}}^{\downarrow X}(x,\bfk)\bigg|-\frac{1}{2};x P^{+},\bfk\bigg\rangle \bigg] \, ,
	\end{eqnarray}
	where
	\begin{eqnarray}
	\psi_{+\frac{1}{2}}^{\downarrow X}(x,\bfk) &=& \frac{(k^{1}-ia k^{2})}{x} \, \varphi_{X} \, , \nonumber \\
	\psi_{-\frac{1}{2}}^{\downarrow X}(x,\bfk) &=& \bigg(M_{X}+\frac{m_{q}}{x}\bigg) \, \varphi_{X} \, .
	\label{WFSNH}
	\end{eqnarray}
	The scalar part $\varphi$ has a form
	\begin{eqnarray}
	\varphi_{X}=\varphi_{X}(x,\bfk)=\frac{\frac{g}{\sqrt{1-x}}}{M^{2}_{X}-\frac{\bfk^{2}+m^{2}_{q}}{x}-\frac{\bfk^{2}+\lambda^{2}_{n}}{1-x}} \, .	\label{scalar}
	\end{eqnarray}
	\section{Numerical Parameters \label{secnumpar}}
    The input parameters used in our calculations are the masses of baryons $M_{X}$, mass of a quark $m_{q}$ and mass of a spectator diquark $\lambda_{n}$. $\Lambda_{X}$ represents the ultraviolet cut off on $\bfk$ which can be evaluated using the fact that $F_{1}(0)=1$ i.e. $\int H_{X}^{q}(x,0,0) dx=1$ such that $\Lambda_{X}>> M_{X}$. Following Ref. \cite{Brodsky:2002cx} for proton, Particle Data Group and Ref. \cite{Zhang:2016qqg} for other particles, the value of the masses of baryons, quarks and diquarks content are listed in Table \ref{tab_bmass} and \ref{tab_qmass}.  For the sake of simplicity, we have considered $m_{u}=m_{d}$ and have represented the contrast of these light quarks with strange quark, $s$. All the plots presented in this paper are in the units of $\frac{g^{2}}{4 \pi}$.
	\begin{table}[h]
		\centering
		\begin{tabular}{|c|c|c|c|c|c|}
			\hline
			$\text{Particle $(X)$}  $~~&~~$ p  $~~&~~$  \Lambda $~~&~~$ \Sigma^{+} $~~&~~$ \Sigma^{o} $~~&~~$ \Xi^{o} $ \\
			\hline
			$\text{Mass, $M_{X}$ (GeV)} $~~&~~$ 0.938 $~~&~~$ 1.115 $~~&~~$ 1.189 $~~&~~$ 1.192 $~~&~~$ 1.314 $ \\
			\hline
		\end{tabular}
		\caption{Masses of members of octet baryons used in the present calculations.}
		\label{tab_bmass} 
	\end{table}
	\begin{table}[h]
		\centering
		\begin{tabular}{|c|c|c|c|c|c|}
			\hline
			$\text{Quark content}  $~~&~~$ u/d  $~~&~~$  s $~~&~~$ uu/ud $~~&~~$ us/ds $~~&~~$ ss $ \\
			\hline
			$\text{Mass (GeV)} $~~&~~$ 0.33 $~~&~~$ 0.48 $~~&~~$ 0.80 $~~&~~$ 0.95 $~~&~~$ 1.10 $ \\
			\hline
		\end{tabular}
		\caption{Masses of quarks and their combinations used in the present calculations.}
		\label{tab_qmass} 
	\end{table}

	\section{Generalized Parton Distributions \label{secgpd}}
	The matrix elements of the bilinear vector currents give the GPDs, $H_{X}^{q}(x,0,\Dp)$ and $E_{X}^{q}(x,0,\Dp)$ which are linked to an unpolarized hadron state $|P\rangle$ \cite{Meissner:2009ww} and are expressed as
	\begin{eqnarray}
	\frac{1}{2}\int \frac{dy^{-}}{2 \pi}&e^{ix P^{+}y^{-}}& \bigg\langle P^{\prime}\bigg|\bar{\psi}\bigg(\frac{-y}{2}\bigg) \,\gamma^{+} \,\psi\bigg(\frac{y}{2}\bigg)\bigg|P\bigg\rangle \bigg|_{y^{+}=0, \bf{y_{\perp}}=0}  \nonumber \\
	&=&\frac{1}{2 \bar{P^+}}\bar{u}(P^\prime)\bigg[H_{X}^{q}(x,0,\Dp) \,\gamma^{+}+E_{X}^{q}(x,0,\Dp) \,\frac{i \sigma^{+\alpha}(-\Delta_{\alpha})}{2M}\bigg]u(P).
	\end{eqnarray} 
	Here, $\bar{u}(P^\prime)$ and ${u}(P)$ are the light-cone spinors of the hadron and $\bar{P}$ represents the average momentum of the initial and final state. Quark fields at two different positions are denoted by $\bar{\psi}(\frac{-y}{2})$ and $\bar{\psi}(\frac{y}{2})$. By utilizing the expressions of a hadron state, quark field and spinors, one can write these off-forward matrix elements in terms of the overlap form of wave functions. As we are dealing with zero skewness distributions, the only relevant kinematical region left is $x\in[0,1]$ which corresponds to the DGLAP (Dokshitzer-Gribov-Lipatov-Altarelli-Parisi-like) region where emission and then reabsorption of a quark takes place.

	For a hadron target comprising of a fermion and scalar spectator, the contributions can be expressed in terms of the two particle light-front wave functions as 
	\begin{eqnarray}
	H_{X(2\rightarrow2)}^{q}(x,0,\Dp) = \int\frac{d^{2}\bfk}{2(2\pi)^3} \bigg[\psi^{\uparrow X \ast}_{+\frac{1}{2}}(x,\bfkpr) \, \psi^{\uparrow X}_{+\frac{1}{2}}(x,\bfk)+ \psi^{\uparrow X \ast}_{-\frac{1}{2}}(x,\bfkpr) \, \psi^{\uparrow X}_{-\frac{1}{2}}(x,\bfk)\bigg], \label{LFWFH} \\
	\frac{(\Delta^{1}-i \Delta^{2})}{2M} E_{X(2\rightarrow2)}^{q}(x,0,\Dp) =  \int\frac{d^{2}\bfk}{2(2\pi)^3} \bigg[\psi^{\uparrow X \ast}_{+\frac{1}{2}}(x,\bfkpr) \, \psi^{\downarrow X}_{+\frac{1}{2}}(x,\bfk)+ \psi^{\uparrow X \ast}_{-\frac{1}{2}}(x,\bfkpr) \, \psi^{\downarrow X}_{-\frac{1}{2}}(x,\bfk)\bigg],
	\label{LFWFE}
	\end{eqnarray}
	where $\bfkpr$ is the final momentum carried by an active quark and it's relation with initial momentum carried by a quark $(\bfk)$ and the momentum transferred $(\Dp)$ is expressed as
	\begin{equation}
		\bfkpr=\bfk-(1-x)\Dp. \nonumber
	\end{equation}
	For the spectator diquark, it is expressed as
	\begin{equation}
		\bfkpr=\bfk+x\Dp. \nonumber
	\end{equation}
	 \begin{figure*}
		\centering
		\begin{minipage}[c]{0.98\textwidth}
			\includegraphics[width=7.5cm]{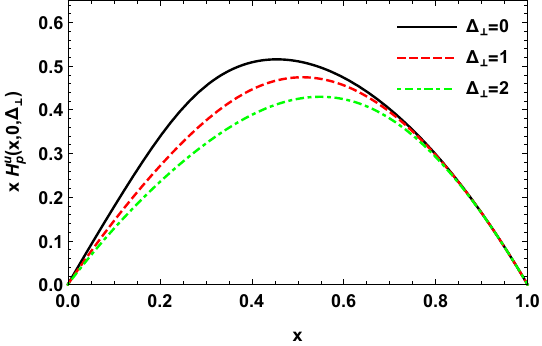}
			\hspace{0.05cm}
		\end{minipage}
		\caption{\label{fig1d2Hp} (Color online) Generalized parton distribution of an unpolarized $u$ quark in proton as a function of $x$ for different values of $\Dp$.}
	\end{figure*}
		\begin{figure*}
		\centering
		\begin{minipage}[c]{0.98\textwidth}
			(a)\includegraphics[width=7.5cm]{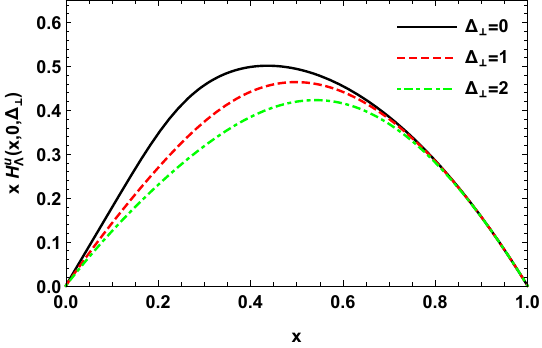}
			\hspace{0.03cm}
			(b)\includegraphics[width=7.5cm]{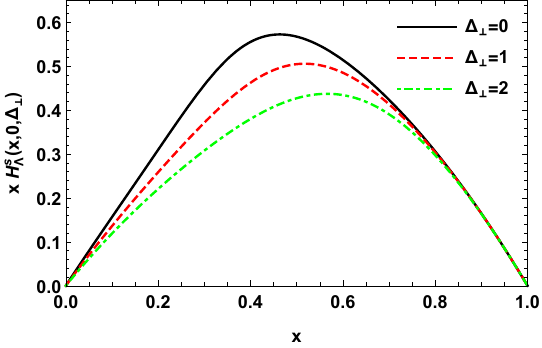}
			\hspace{0.03cm}
			(c)\includegraphics[width=7.5cm]{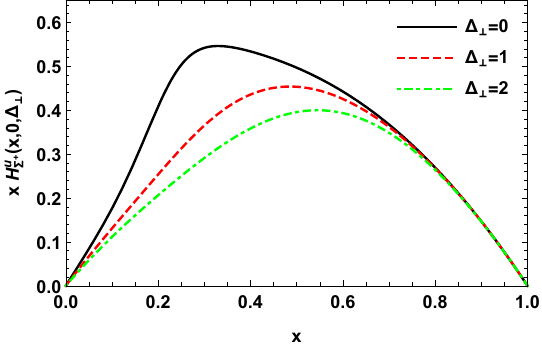}
			\hspace{0.03cm}
			(d)\includegraphics[width=7.5cm]{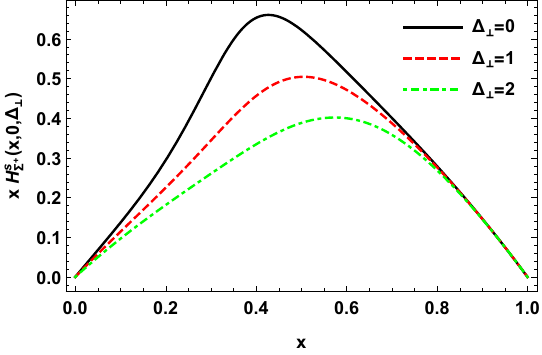}
			\hspace{0.03cm}
			(e)\includegraphics[width=7.5cm]{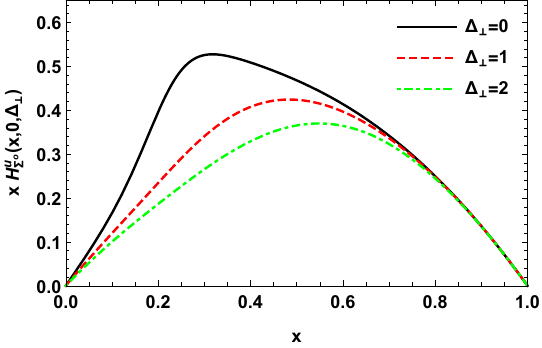}
			\hspace{0.03cm}
			(f)\includegraphics[width=7.5cm]{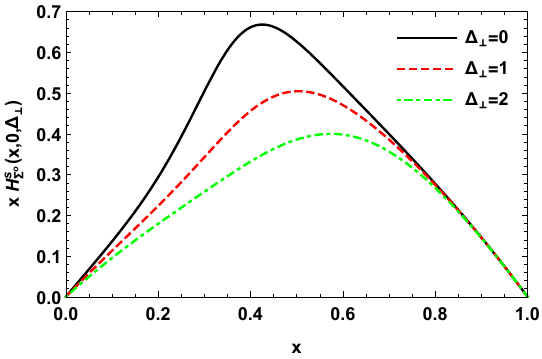}
			\hspace{0.03cm}			
		\end{minipage}
		\caption{\label{fig2d2Hs} (Color online) Generalized parton distributions of unpolarized quarks for baryons with single $s$ content as a function of $x$ for different values of $\Dp$. The left and right column correspond to $u$ and $s$ quarks sequentially.}
	\end{figure*} 
	On substituting the expressions of light-front wave functions from Eqs. (\ref{WFSPH}) in Eq. (\ref{LFWFH}), we can obtain an explicit expression of the GPD $H_{X}^{q}(x,0,\Dp)$ and it is expressed as
	\begin{eqnarray}
	H_{X}^{q}(x,0,\Dp)=\frac{g^{2}}{2(2\pi)^{3}}[I_{1X}+I_{2X}+(C_{X}+2 \,(x \, M_{X} +m_{q})^{2}) \, I_{3X}]\frac{(1-x)}{2} \, , \label{ExpH}
	\end{eqnarray}
	with
	\begin{eqnarray}
	I_{1X} &=& \int \frac{d^{2}\bfk}{L_{1X}}=\pi log \bigg(\frac{\Lambda^{2}_{X}}{\mathcal{M}_{X}}\bigg) \, , \nonumber \\
	I_{2X} &=& \int \frac{d^{2}\bfk}{L_{2X}}=\pi log \bigg(\frac{\Lambda^{2}_{X}}{\mathcal{M}_{X}+\Dp^{2} (1-x)^{2}}\bigg) \, , \nonumber \\
	I_{3X} &=& \pi \int_{0}^{1} \frac{d\alpha}{D_{X}}\,,
	\end{eqnarray}
	where
	\begin{eqnarray}
	L_{1X} &=&\bfk^{2}+\mathcal{M}_{X} \, , \nonumber \\
	L_{2X} &=&{{\bf k}_{\perp}^{\prime 2}}+\mathcal{M}_{X} \, , \nonumber \\
	C_{X} &=& -2 \mathcal{M}_{X}-(1-x)^{2}\Dp^{2} \, , \nonumber \\
	D_{X} &=& \alpha (1-\alpha)(1-x)^{2}\Dp^{2}+\mathcal{M}_{X} \, , \nonumber \\
	\mathcal{M}_{X} &=& m_{q}^{2}(1-x)+\lambda^{2}_{n}x-M^{2}_{X}x(1-x) \, . 
	\end{eqnarray}

	 \begin{figure*}
		\centering
		\begin{minipage}[c]{0.98\textwidth}
		(a)\includegraphics[width=7.5cm]{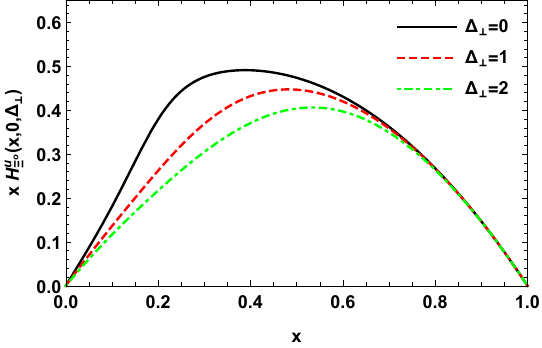}
		\hspace{0.03cm}
		(b)\includegraphics[width=7.5cm]{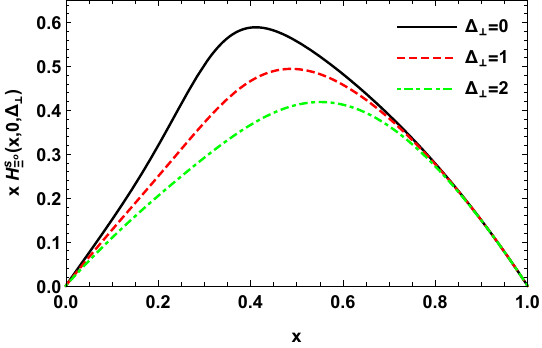}
		\hspace{0.03cm}
		\end{minipage}
		\caption{\label{fig3d2Hss} (Color online) Generalized parton distributions of unpolarized quarks for baryon with double $s$ content as a function of $x$ at different values of $\Dp$. The left and right column correspond to $u$ and $s$ quarks sequentially.}
	\end{figure*}
	Likewise, on substituting the expressions of light-front wave functions from Eqs. (\ref{WFSPH}) and (\ref{WFSNH}) in Eq. (\ref{LFWFE}), we can obtain an explicit expression of the  GPD $E_{X}^{q}(x,0,\Dp)$ which is expressed as
	\begin{eqnarray}
	E_{X}^{q}(x,0,\Dp)=\frac{g^{2}}{2(2\pi)^{3}} 2M_{X} \, (1-x)^{2} I_{3X} \, . \label{ExpE}
	\end{eqnarray}
   
	The distributions of GPDs are basically the interference of amplitudes and  interpret different quantum fluctuations of a baryon. The distribution of a GPD $H_{X}^{q}(x,0,\Dp)$  corresponding to an active $u$ quark for proton is represented in the Fig. \ref{fig1d2Hp}. In Figs. \ref{fig2d2Hs} and \ref{fig3d2Hss}, the left panels represent the distributions corresponding to the active quark being a $u$ quark and the right panel for the active quark being a $s$ quark.  In Fig. \ref{fig2d2Hs} the baryons with a single $s$ quark are presented ($\Sigma$ and $\Lambda$) and in Fig. \ref{fig3d2Hss}, the baryon with double $s$ content are presented ($\Xi$). Moving from a baryon having no strange quark to a baryon having a single or double strange quark(s), it has been observed that the likeliness of a $u$ quark to take away the fraction of a longitudinal momentum from its parent baryon decreases as the peak is shifted towards the lower values of $x$. This may be attributed to the fact that the momentum carried by an object is proportional to its mass. As we move from $p$ (carrying no $s$ quark) to $\Sigma^{+}, \Sigma^{o}$, $\Lambda$ and $\Xi^{o}$ (carrying a single or double $s$ quark(s)), a diquark becomes heavier and can carry large fraction of longitudinal momentum than an active quark, $u$. On an average, there is a decrement of about $8\%$ of the ability of $u$ quark to carry a fraction of longitudinal momentum from its parent baryon. However, only a minute difference is observed while moving from $\Sigma^{+}$ $(\Sigma^{o}$ or $\Lambda)$ to $\Xi^{o}$ in case of a distribution of unpolarized  quarks in unpolarized baryon. The distributions corresponding to an unpolarized active $s$ quark in unpolarized baryon are represented in the right panel of Figs. \ref{fig2d2Hs} and \ref{fig3d2Hss}. Since $p$ does not have a strange content, comparison will be considered between a baryon having a single  and double $s$ quark content. A decrement in the capacity of $s$ quark to take away the fraction of longitudinal momentum from its parent baryon is also observed here as we go from a light diquark $(uu$ or $ud)$ in case of $\Sigma^{+}$ $(\Sigma^{o}$ or $\Lambda)$ to a heavy diquark $(us)$ of $\Xi^{o}$ and this decrement is about $3\%$. The amplitude of distributions for $s$ as an active quark is observed to be more than an active $u$ quark which signifies more constructive interference occurring among the different quantum fluctuations of a baryon for an active $s$ quark. 

\begin{figure*}
		\centering
		\begin{minipage}[c]{0.98\textwidth}
			(a)\includegraphics[width=7.5cm]{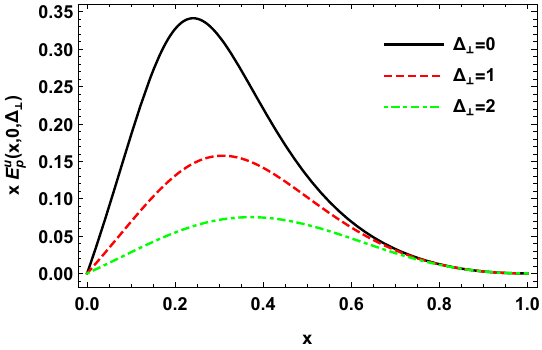}
			\hspace{0.05cm}
		\end{minipage}
		\caption{\label{fig4d2Ep} (Color online) Generalized parton distribution of a transversely polarized $u$ quark for a proton as a function of $x$ for different values of $\Dp$.}
	\end{figure*}

\begin{figure*}
		\centering
		\begin{minipage}[c]{0.98\textwidth}
			(a)\includegraphics[width=7.5cm]{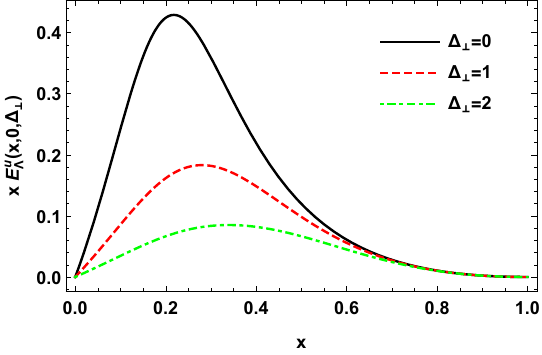}
			\hspace{0.03cm}
			(b)\includegraphics[width=7.5cm]{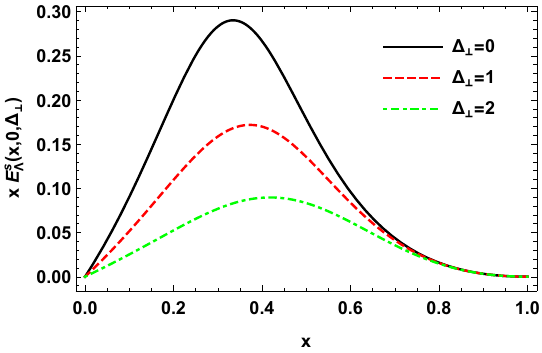}
			\hspace{0.03cm}
			(c)\includegraphics[width=7.5cm]{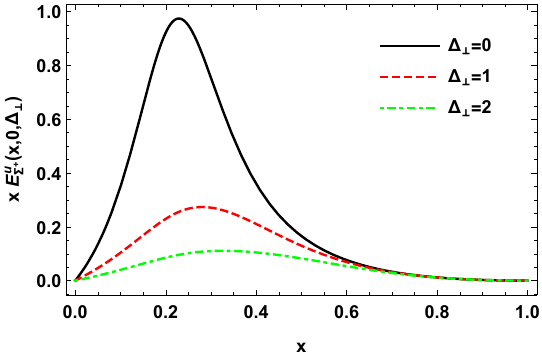}
			\hspace{0.03cm}
			(d)\includegraphics[width=7.5cm]{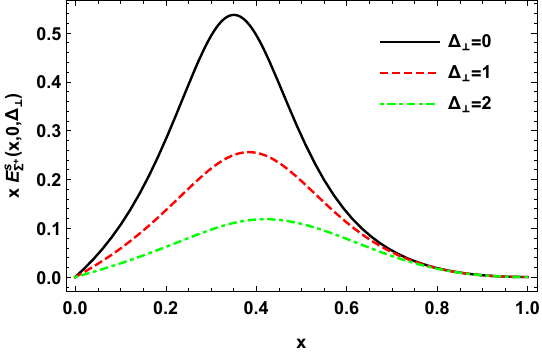}
			\hspace{0.03cm}
			(e)\includegraphics[width=7.5cm]{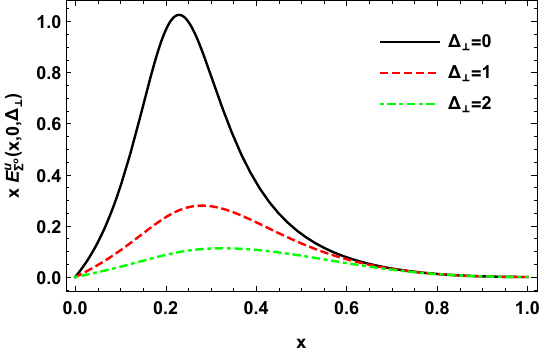}
			\hspace{0.03cm}
			(f)\includegraphics[width=7.5cm]{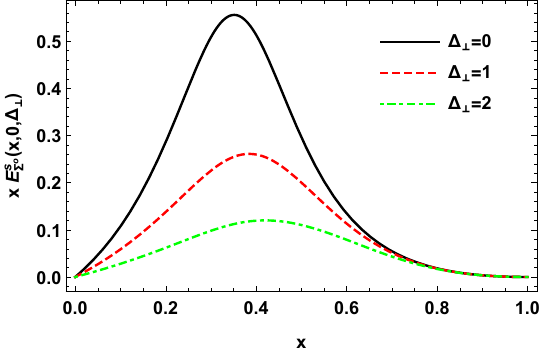}
			\hspace{0.03cm}
		\end{minipage}
		\caption{\label{fig5d2Es} (Color online) Generalized parton distributions of transversely polarized quarks for baryons with single $s$ content as a function of $x$ for different values of $\Dp$. The left and right column correspond to $u$ and $s$ quarks sequentially.}
	\end{figure*}

\begin{figure*}
		\centering
		\begin{minipage}[c]{0.98\textwidth}
			(a)\includegraphics[width=7.5cm]{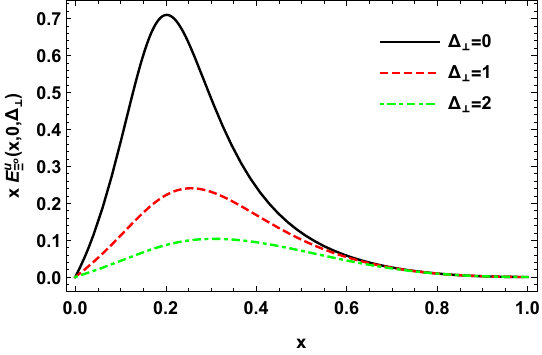}
			\hspace{0.03cm}
			(b)\includegraphics[width=7.5cm]{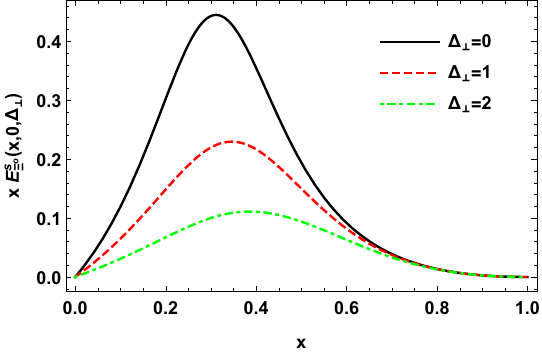}
			\hspace{0.03cm}	
		\end{minipage}
		\caption{\label{fig6d2Ess} (Color online) Generalized parton distributions of transversely polarized quarks for baryon with double $s$ content as a function of $x$ for different values of $\Dp$. The left and right column correspond to $u$ and $s$ quarks sequentially.}
	\end{figure*}

	The distributions corresponding to transversely polarized active $u$ quark are represented in Fig. \ref{fig4d2Ep} for proton. Whereas the left panels of Figs. \ref{fig5d2Es} and \ref{fig6d2Ess} correspond to the strangeness carrying baryons. Discussing in the same way from baryon having no strange quark to a baryon having a single or double strange quark(s) as discussed for unpolarized quarks, it has been perceived that the ability of $u$ quark to carry a fraction of longitudinal momentum from its parent baryon decreases but this decrement is only about $1.3\%$ from $p$ to $\Sigma^{+}$ $(\Sigma^{o}$ or $\Lambda)$ and just $1\%$ from $\Sigma^{+}$ $(\Sigma^{o}$ or $\Lambda)$ to $\Xi^{o}$. The distributions of  a transversely polarized active $s$ quark are also represented in the right panel of Figs. \ref{fig5d2Es} and \ref{fig6d2Ess} which show that the decrease in the capacity of $s$ quark to take away the fraction of longitudinal momentum from its parent baryon is found to be $3\%$ as we jump from $\Sigma^{+}$ $(\Sigma^{o}$ or $\Lambda)$ to $\Xi^{o}$. The amplitudes of the distributions for both $u$ and $s$ quark are found to increase in going from $p$ to  $\Sigma^{+}$ $(\Sigma^{o}$ or $\Lambda)$. This corresponds to more a constructive interference among different quantum fluctuations inside a baryon having single massive quark. However, the amplitude of $\Xi^{o}$ having two $s$ quarks decreases which is due to the dominance of the diquark mass. On the other hand, the amplitude of the distributions for $u$ as an active quark is observed to be more than when the active quark is a $s$ quark. This clearly signifies more constructive interference among the different quantum fluctuations of a baryon for an active $u$ quark than $s$ quark.   \par

	In a nut shell, almost same kind of trend is followed by the distribution of unpolarized  quarks in unpolarized hadron, $H_{X}^{q}(x,0,\Dp)$ and a transversely polarized quarks in unpolarized hadron, $E_{X}^{q}(x,0,\Dp)$ as we jump from $p$ (carrying no $s$ quark) to $\Sigma^{+}$ ($\Sigma^{o}$ or $\Lambda$) (carrying a single $s$ quark) and then to $\Xi^{o}$ (carrying double $s$ quarks). However, the distributions of $E_{X}^{q}(x,0,\Dp)$ are much narrower and peaked at lower value of $x$ for all particles than $H_{X}^{q}(x,0,\Dp)$. It implies that the transversely polarized quark in an unpolarized baryon can take away less fraction of the longitudinal momentum from its parent baryon as compared to  the unpolarized quark in an unpolarized baryon. It holds for both kind of $u$ and $s$ quarks with the only difference being in the amplitude of the distributions which comes out be more for an unpolarized active $s$ quark and less for a transversely polarized active $s$ quark than an active $u$ quark.
	
    For the distributions of unpolarized quarks, on increasing the value of a longitudinal momentum transferred $\Dp$ from zero to different finite values, the ability of an active quark to take away the fraction of the longitudinal momentum increases as the peaks of the distributions move towards higher value of $x$, however, with the amplitude of distributions decreasing. This decrement for amplitudes in case of an active $s$ quark is more than an active $u$ quark which implies that there is more destructive interference of the quantum fluctuations for an unpolarized active $s$ quark inside the baryon on change of one unit of longitudinal momentum transferred. By observing the distributions of transversely polarized quarks, we can interpret that the constructive interference taking place among different quantum fluctuations is more significant for $s$ quarks than $u$ quark as the amplitude decreases more for $u$ quark than $s$ quark.    Despite this, at large values of $x$, distributions corresponding to both unpolarized as well as transversely polarized quarks becomes independent of $\Dp$ irrespective of the flavor of an active quark.
    
	\section{Impact Parameter Dependent Parton Distributions \label{secipdpdf}}
	\begin{figure*}
		\centering
		\begin{minipage}[c]{0.98\textwidth}
			(a)\includegraphics[width=7.5cm]{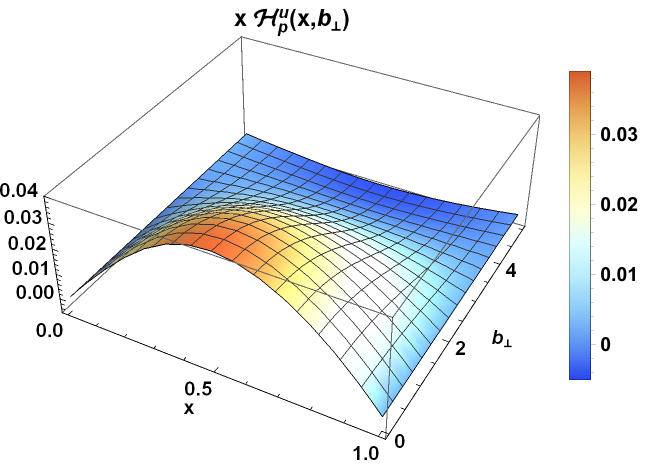}
			\hspace{0.05cm}
		\end{minipage}
		\caption{\label{fig7d3Hp} (Color online) Impact parameter dependent parton distribution of an unpolarized $u$ quark in proton as a function of $x$ and $|\bfb|$.}
	\end{figure*}
	\begin{figure*}
		\centering
		\begin{minipage}[c]{0.98\textwidth}
			(a)\includegraphics[width=7.5cm,clip]{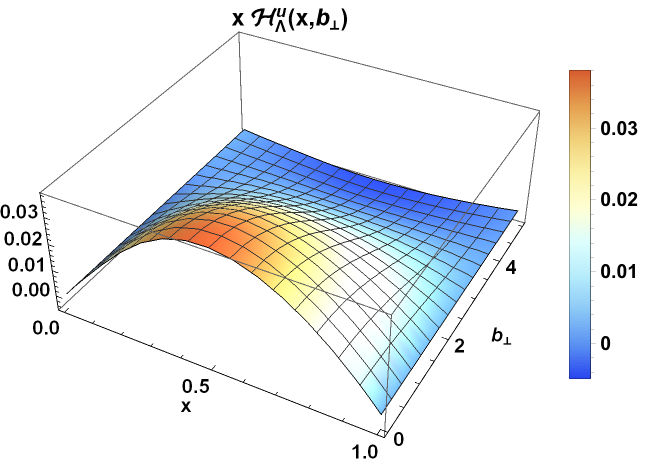}
			\hspace{0.02cm}
			(b)\includegraphics[width=7.5cm,clip]{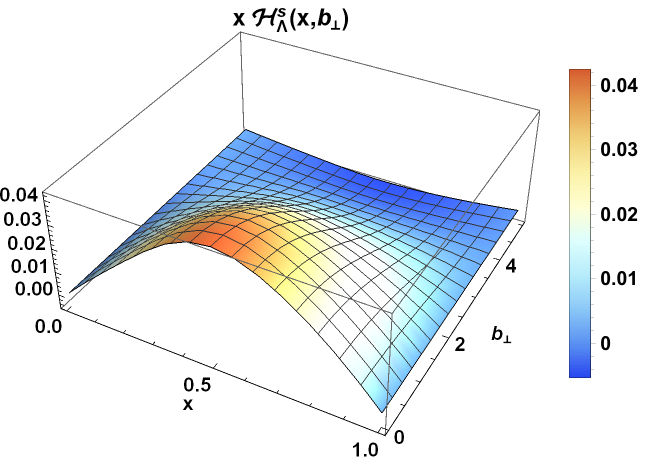}
			\hspace{0.02cm}
			(c)\includegraphics[width=7.5cm,clip]{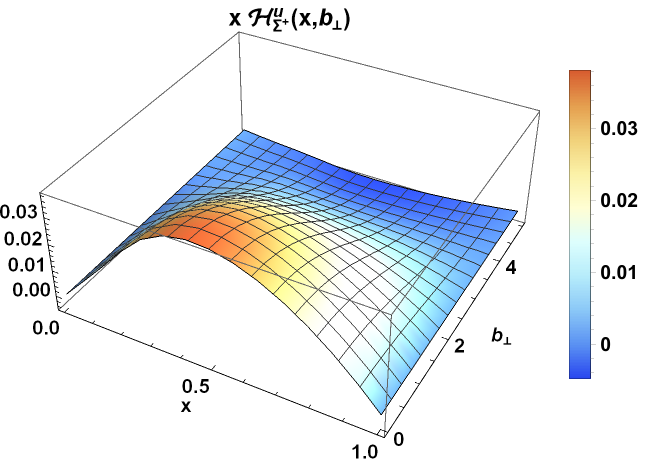}
			\hspace{0.02cm}
			(d)\includegraphics[width=7.5cm,clip]{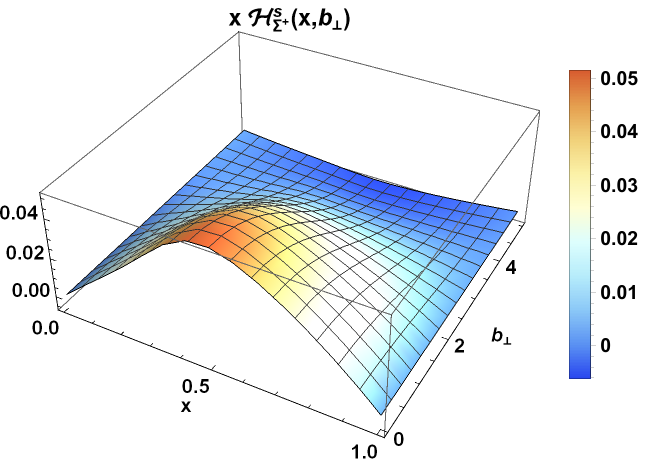}
			\hspace{0.02cm}
			(e)\includegraphics[width=7.5cm,clip]{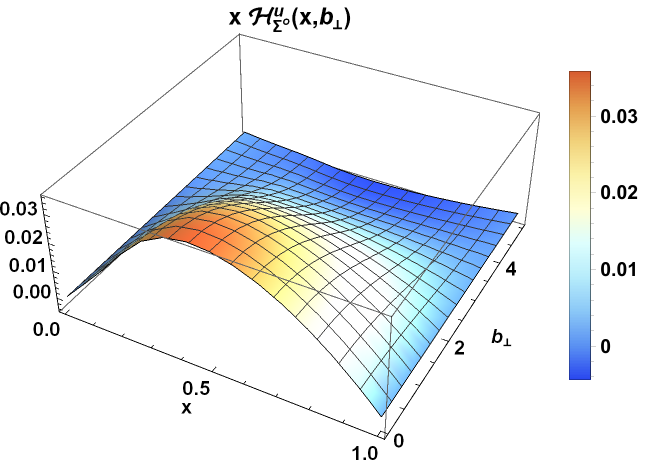}
			\hspace{0.02cm}
			(f)\includegraphics[width=7.5cm,clip]{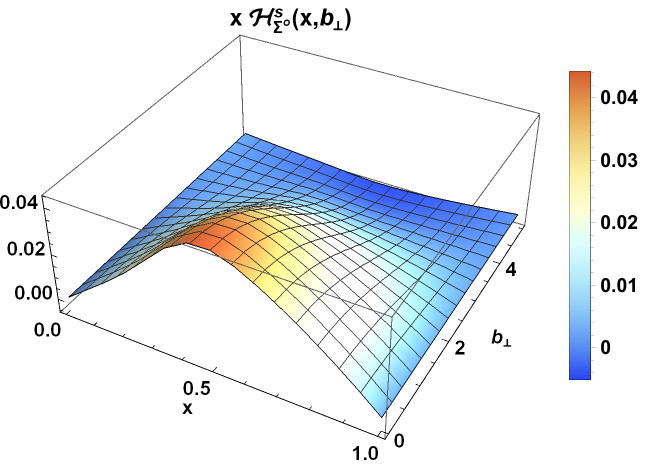}
			\hspace{0.02cm} \\
		\end{minipage}
		\caption{\label{fig8d3Hs} (Color online) Impact parameter dependent parton distributions of unpolarized quarks for baryons with single $s$ content as a function of $x$ and $|\bfb|$. The left and right column correspond to $u$ and $s$ quarks sequentially.}
	\end{figure*}
	\begin{figure*}
		\centering
		\begin{minipage}[c]{0.98\textwidth}
		(a)\includegraphics[width=7.5cm,clip]{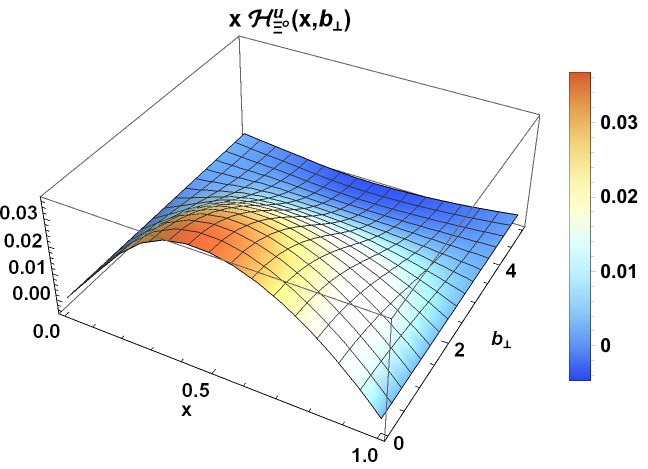}
		\hspace{0.02cm}
		(b)\includegraphics[width=7.5cm,clip]{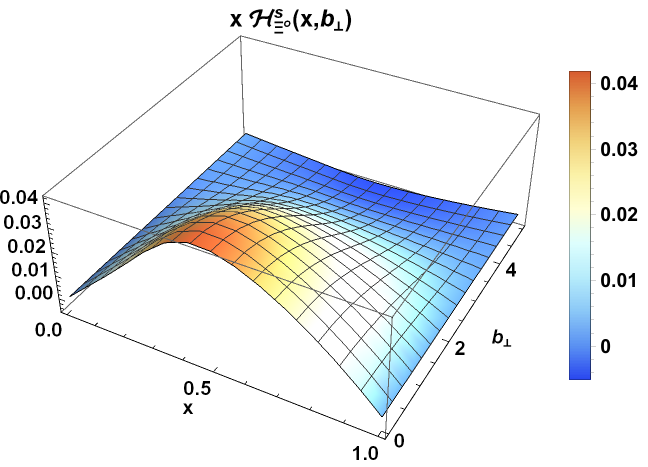}
		\hspace{0.02cm} \\
		\end{minipage}
		\caption{\label{fig9d3Hss} (Color online) Impact parameter dependent parton distributions of unpolarized quarks for baryon with double $s$ content as a function of $x$ and $|\bfb|$. The left and right column correspond to $u$ and $s$ quarks sequentially.}
	\end{figure*}
   IPDPDFs lead us to the transverse plane and one can obtain the distributions in impact parameter space on operating the Fourier transformation on GPDs in momentum space \cite{Burkardt:2000wc}. Magnitude of the impact parameter $|\bfb|$ is analogous to the length and can be described as the perpendicular distance between an active quark and the centre of momentum of a baryon. Hence, it represents the transverse position of an active quark with respect to a baryon. Such distributions \cite{Burkardt:2002hr} are defined as
	\begin{eqnarray}
	\mathcal{H}(x,0,\bfb)=\frac{1}{(2\pi)^2} \int d^{2}\Dp e^{-i \bfb\cdot\Dp}H_{X}^{q}(x,0,\Dp), \\
	\mathcal{E}(x,0,\bfb)=\frac{1}{(2\pi)^2} \int d^{2}\Dp e^{-i \bfb\cdot\Dp}E_{X}^{q}(x,0,\Dp).
	\end{eqnarray}
	One can write these expressions in terms of the first kind of Bessel function of order zero as
	\begin{eqnarray}
	\mathcal{H}(x,0,\bfb)=\frac{1}{2\pi} \int d^{2}\Delta \, J_{o}(b\Delta) \, H_{X}^{q}(x,0,\Dp), \\
	\mathcal{E}(x,0,\bfb)=\frac{1}{2\pi} \int d^{2}\Delta \, J_{o}(b\Delta) \, E_{X}^{q}(x,0,\Dp),
	\end{eqnarray}
	where the expressions for $H_{X}^{q}(x,0,\Dp)$ and $E_{X}^{q}(x,0,\Dp)$ can be taken from Eqs. $(\ref{ExpH})$ and $(\ref{ExpE})$. \par
	\begin{figure*}
		\centering
		\begin{minipage}[c]{0.98\textwidth}
			\includegraphics[width=7.5cm]{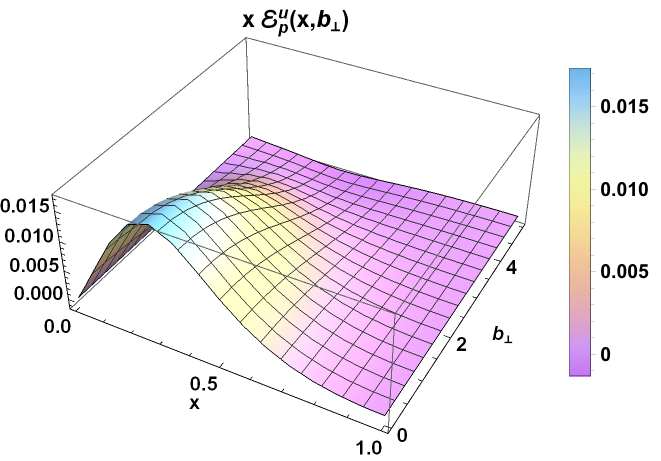}
			\hspace{0.05cm}
		\end{minipage}
		\caption{\label{fig10d3Ep} (Color online) Impact parameter dependent parton distribution of a transversely polarized $u$ quark in proton as a function of $x$ and $|\bfb|$.}
	\end{figure*}
\begin{figure*}
	\centering
	\begin{minipage}[c]{0.98\textwidth}
		(a)\includegraphics[width=7.5cm]{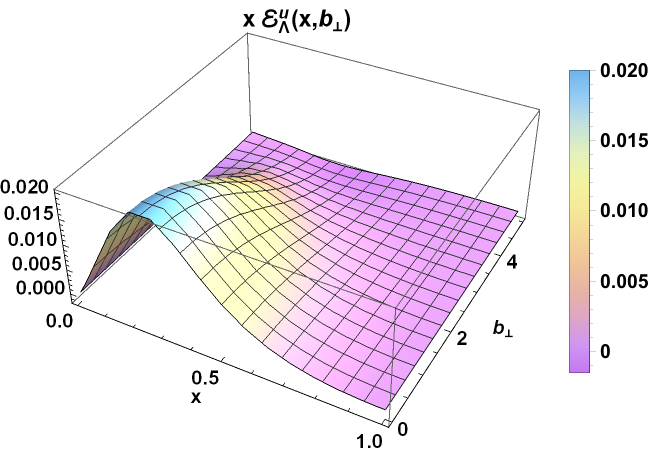}
		\hspace{0.02cm}
		(b)\includegraphics[width=7.5cm]{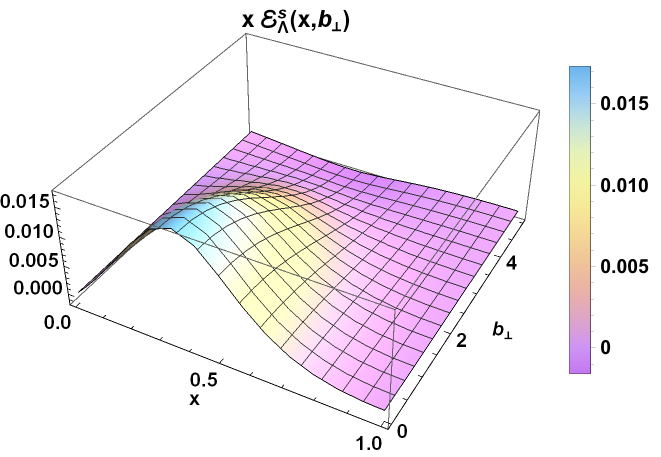}
		\hspace{0.02cm}
		(c)\includegraphics[width=7.5cm]{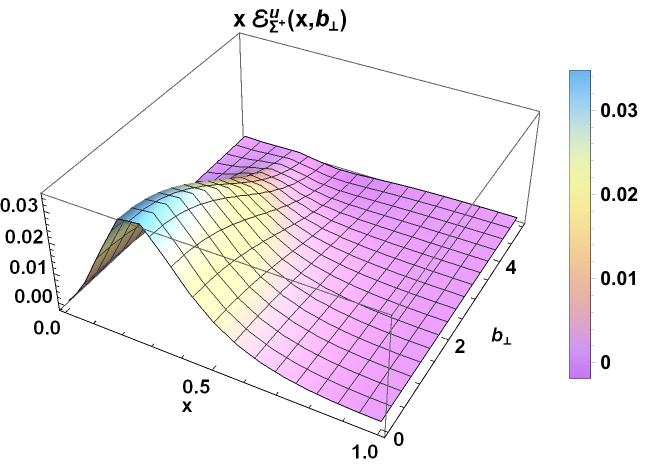}
		\hspace{0.02cm}
		(d)\includegraphics[width=7.5cm]{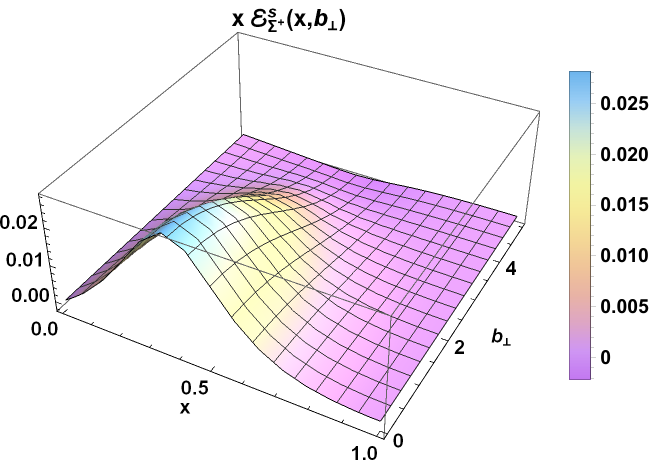}
		\hspace{0.02cm}
		(e)\includegraphics[width=7.5cm]{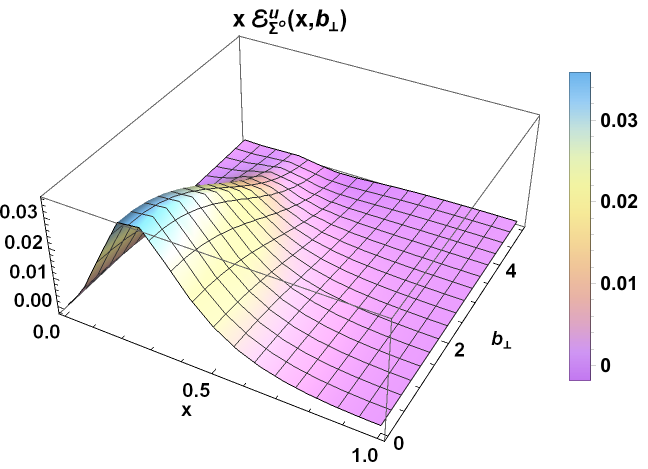}
		\hspace{0.02cm}
		(f)\includegraphics[width=7.5cm]{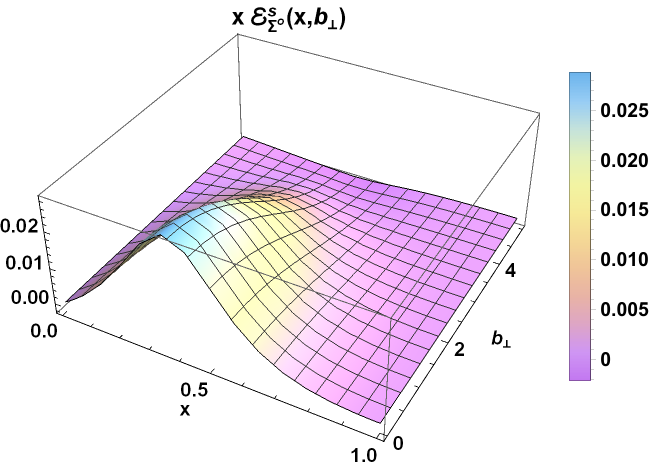}
		\hspace{0.02cm} \\
	\end{minipage}
	\caption{\label{fig11d3Es} (Color online) Impact parameter dependent parton distributions of transversely polarized quarks for baryons with single $s$ content as a function of $x$ and $|\bfb|$. The left and right column correspond to $u$ and $d$ quarks sequentially.}
\end{figure*}
\begin{figure*}
	\centering
	\begin{minipage}[c]{0.98\textwidth}
		(a)\includegraphics[width=7.5cm]{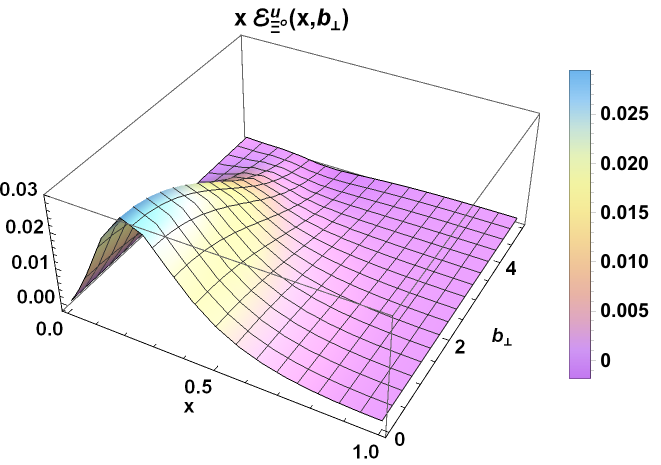}
		\hspace{0.02cm}
		(b)\includegraphics[width=7.5cm]{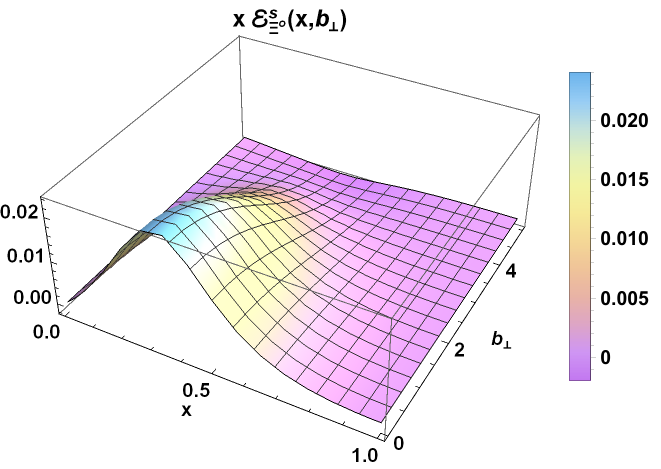}
		\hspace{0.02cm} \\
	\end{minipage}
	\caption{\label{fig12d3Ess} (Color online) Impact parameter dependent parton distributions of transversely polarized quarks for baryon with double $s$ content as a function of x and $|\bfb|$. The left and right column correspond to $u$ and $d$ quarks sequentially.}
\end{figure*}
	The variations in distributions of unpolarized quarks with respect to longitudinal momentum fraction carried by a quark and its impact parameter are demonstrated in Figs. \ref{fig7d3Hp}, \ref{fig8d3Hs} and \ref{fig9d3Hss}. While moving from $p$ to $\Sigma^{+}$ ($\Sigma^{o}$ or $\Lambda$) and then to $\Xi^{o}$, the peak is displaced towards the lower values of longitudinal momentum fraction and also becomes narrower. The causes of these observed changes are same as mentioned for the distributions in momentum space and these changes reassure our results. The juxtapose plots of $u$ and $s$ as an active quarks are presented in Figs. \ref{fig8d3Hs} and \ref{fig9d3Hss}. They show that the maxima for $s$ quarks lies at higher values of $x$ as compared to those for $u$ quarks implying massive object carrying more fraction of longitudinal momentum. The over all small amplitude signifies less constructive interference among quantum fluctuations but in general the variations recorded while moving from $p$ to $\Sigma^{+}$ ($\Sigma^{o}$ or $\Lambda$) and then to $\Xi^{o}$ are same for $u $ and $s$ quarks.  
	
	The distributions corresponding to a transversely polarized quarks for proton, baryons with single strangeness content and a baryon with double strangeness content are presented in Figs. \ref{fig10d3Ep}, \ref{fig11d3Es} and \ref{fig12d3Ess} respectively. On moving from $p$ to $\Sigma^{+}$ ($\Sigma^{o}$ or $\Lambda$) and then to $\Xi^{o}$, the change in the amplitude of the distributions corresponding to both $u$ and $s$ quarks is  marginal, the peak moves to lower values of longitudinal momentum fraction and also become narrower. Here, the increase of amplitude is a consequence of respective heavy diquark mass inside the baryon. The width of the distributions for $s$ quarks is broader and maxima lies at higher value of $x$ than $u$ which implies that $s$ quark always carries more fraction of momentum fraction than the $u$ quark. This is irrespective of it being unpolarized or transversely polarized.
	 
	With an increase in the value of $\bfb$, one of the customary aspects for both  $\mathcal{H}(x,0,\bfb)$ and $\mathcal{E}(x,0,\bfb)$ is the reduction of the width of distributions implying that the distributions are more intense and the quarks are more confined at $\bfb=0$ while having a maximum longitudinal momentum fraction. Another aspect is the movement of a peak to lesser values of $x$. In the context of light-front this clearly implies that smaller the momentum fraction, larger is the kinetic energy carried by the quark. The distributions obtained with the consideration of one-loop quantum fluctuation of Yukawa theory follow the footprints of these customary aspects. It is important to mention here that these aspects are model independent effects of IPDPDFs and have been can discussed in other models as well \cite{Maji:2017ill, Gutsche:2011uj}.
	
	Fall off of the distributions in the direction of increasing $\bfb$ is more rapid for active $s$ quark as compared to that for  $u$ quark and it is clearly visible in the distributions of transversely polarized quarks when compared to those in the case of unpolarized quarks. This rapid fall off for $s$ quark than $u$ represents that there is a little chance of finding an active $u$ quark at higher $\bfb$ whereas likeliness of getting $s$ quark vanishes very early at small $\bfb$.  While comparing the distributions of baryons containing single or double  $s$ quark(s) presented in the left panels of Figs. \ref{fig8d3Hs}, \ref{fig9d3Hss}, \ref{fig11d3Es} and \ref{fig12d3Ess}, it has been observed that the distribution fall off very slowly in baryon having a single $s$ quark which interpret that there is more chance of getting an active $u$ quark at larger $\bfb$ while shifting towards a smaller values of longitudinal momentum fraction carried by it when the diquark is lighter. Similar fashion can be seen in the right panels of Figs. \ref{fig8d3Hs}, \ref{fig9d3Hss}, \ref{fig11d3Es} and \ref{fig12d3Ess} which again implies that the a chance of getting an active $s$ quark with smaller values of longitudinal momentum fraction in the case of single $s$ baryons is more likely as compared to double $s$ baryons. These effects predict that a baryon with double $s$ quarks is more tightly bidden than the baryon with a single $s$ quark.
	
	\section{Charge Distributions in Transverse Coordinate Space and Impact Parameter Space \label{secchargedis}}
		\begin{figure*}
		\centering
		\begin{minipage}[c]{0.98\textwidth}
			\includegraphics[width=7.5cm]{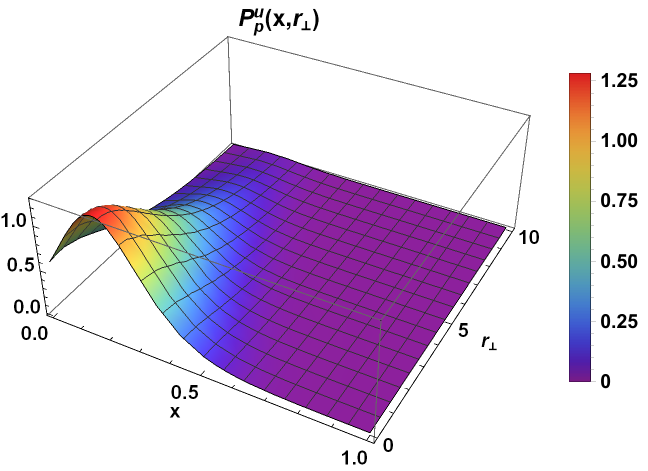}
			\hspace{0.05cm}
		\end{minipage}
		\caption{\label{fig13d3Cordp} (Color online) Charge distribution in coordinate space for $u$ quark in proton.}
	\end{figure*}
	\begin{figure*}
		\centering
		\begin{minipage}[c]{0.98\textwidth}
			(a)\includegraphics[width=7.5cm,clip]{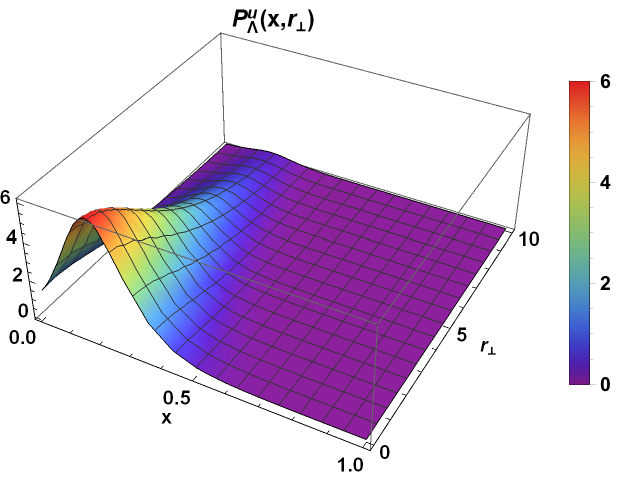}
			\hspace{0.03cm}
			(b)\includegraphics[width=7.5cm,clip]{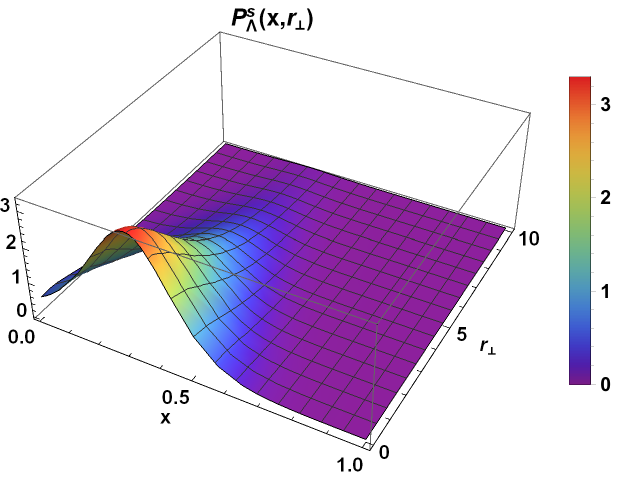}
			\hspace{0.03cm}
			(c)\includegraphics[width=7.5cm,clip]{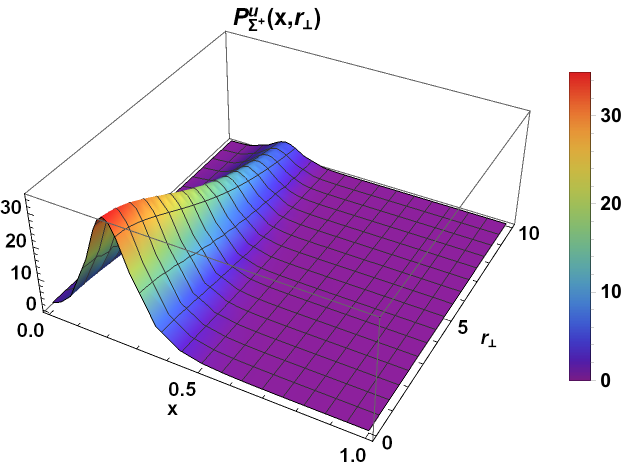}
			\hspace{0.03cm}
			(d)\includegraphics[width=7.5cm,clip]{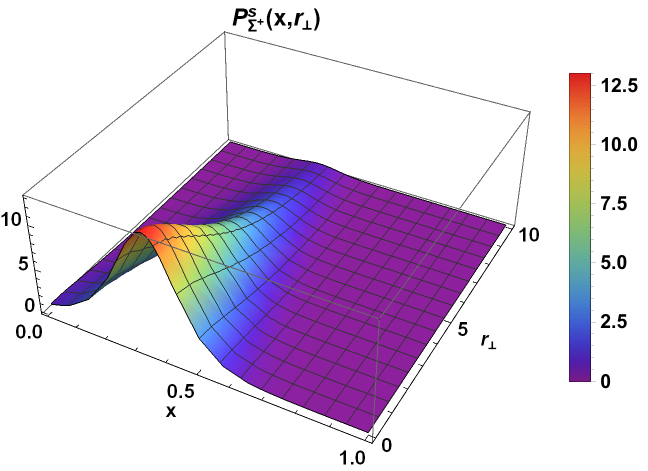}
			\hspace{0.03cm}
			(e)\includegraphics[width=7.5cm,clip]{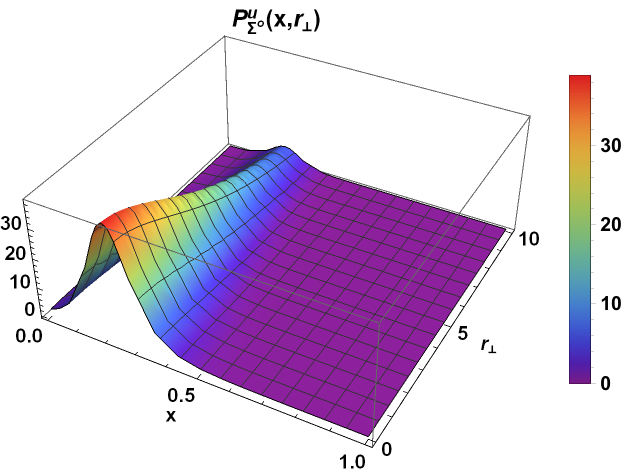}
			\hspace{0.03cm}
			(f)\includegraphics[width=7.5cm,clip]{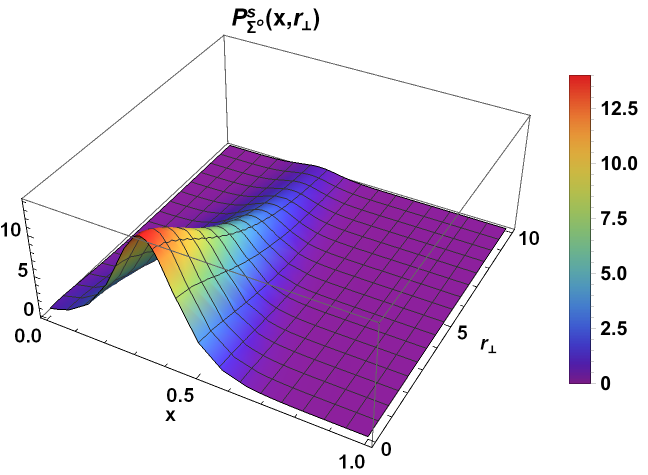}
			\hspace{0.03cm} \\
		\end{minipage}
		\caption{\label{fig14d3Cords} (Color online) Charge distributions in coordinate space for baryons with single $s$ content. The left and right column correspond to $u$ and $s$ quarks sequentially.}
	\end{figure*}
		\begin{figure*}
		\centering
		\begin{minipage}[c]{0.98\textwidth}
			(a)\includegraphics[width=7.5cm,clip]{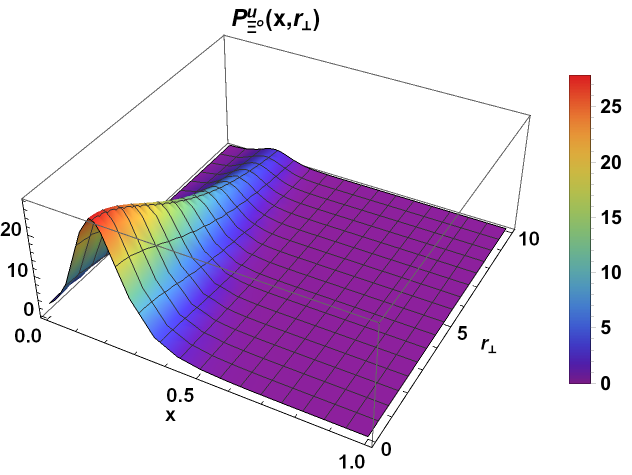}
			\hspace{0.03cm}
			(b)\includegraphics[width=7.5cm,clip]{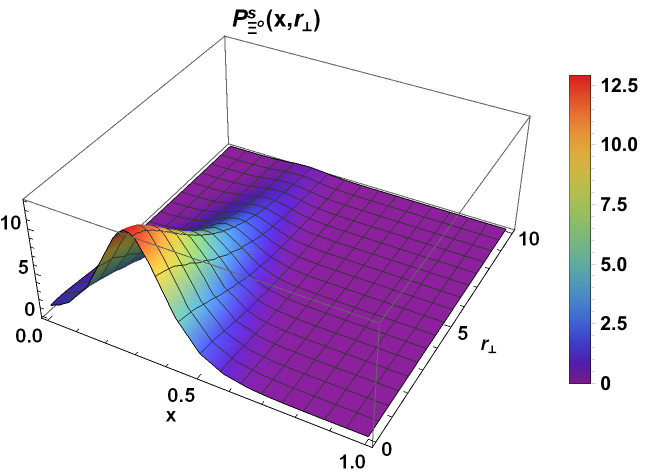}
			\hspace{0.03cm} \\
		\end{minipage}
		\caption{\label{fig15d3Cordss} (Color online) Charge distributions in coordinate space for baryon with double $s$ content. The left and right column correspond to $u$ and $s$ quarks sequentially.}
	\end{figure*}
	In this section, we discuss the charge distributions in transverse coordinate space and impact parameter space to perceive the difference between them. The scalar part of the wave function, mentioned in Eq. $(\ref{scalar})$ and arrived from the Feynman diagram due to the contribution of a spectator propagator \cite{Brodsky:1998hn}, can be generalized by harmonizing its power behavior $j$ as
	\begin{eqnarray}
	\varphi_{X}(x,\bfk)=\frac{g}{\sqrt{1-x}} M^{2j}_{X} x^{-j}\bigg(M^{2}_{X}-\frac{\bfk^{2}+m^{2}_{q}}{x}-\frac{\bfk^{2}+\lambda^{2}_{n}}{1-x}\bigg)^{-j-1}.
	\end{eqnarray}
	The Yukawa theory is designed for $j=0$ and the equation written above has an extra factor of $ M^{2j}_{X} x^{-j} \bigg(M^{2}_{X} - \frac{\bfk^{2}+m^{2}_{q}}{x} -\frac{\bfk^{2}+\lambda^{2}_{n}}{1-x}\bigg)^{-j}$ as compared to Eq. $(\ref{scalar})$. Here, the factor $M^{2j}_{X}$ is considered for the aptness of dimension and the rest part is incorporated from a Lorentz invariant form factor $(k^{2}-m^{2}_{q})$ owing to quark-diquark vertex \cite{Jakob:1997wg}.

	By employing Fourier transformation, one can relate the light-front wave functions in transverse coordinate space and momentum space as
	\begin{eqnarray}
	\psi^{X}(x,\bfk)=\int d^{2}\bfr e^{-i \bfk \cdot \bfr} \tilde{\psi}^{X} (x,\bfr), \nonumber \\
	\tilde{\psi}^{X}(x,\bfr)=\int \frac{d^{2}\bfk}{(2\pi)^{2}} e^{-i \bfk \cdot \bfr}\psi^{X}(x,\bfk). \label{coordspa}
	\end{eqnarray}
	From these equations, we can write
	\begin{eqnarray}
	\int \frac{d^{2}\bfk}{(2\pi)^{2}} \psi^{X \ast} (x,\bfk)  \psi^{X}(x,\bfk)=\int d^{2}\bfr \tilde{\psi}^{X \ast}(x,\bfr) \tilde{\psi}^{X}(x,\bfr).
	\end{eqnarray}
	By using Eqs. (\ref{WFSPH}), (\ref{WFSNH}) and (\ref{coordspa}), the LFWFs for $J^{z}=+\frac{1}{2}$ hadron in coordinate space \cite{Kim:2008ghb} can be expressed as
	\begin{eqnarray} 
	\tilde{\psi}^{\uparrow X}_{+\frac{1}{2}}(x,\bfr)&=&(x \, M_{X}+m_{q}) \frac{\bfr}{\sqrt{\mathcal{M}_{X}}} \frac{g M^{2}_{X}}{4\pi} (1-x)^{\frac{3}{2}} K_{1}\bigg(\bfr \sqrt{\mathcal{M}_{X}}\bigg), \nonumber \\
	\tilde{\psi}^{\uparrow X}_{-\frac{1}{2}}(x,\bfr)&=&-i(r^{1}+i r^{2})  \frac{g M^{2}_{X}}{4\pi} (1-x)^{\frac{3}{2}} K_{0}\bigg(\bfr \sqrt{\mathcal{M}_{X}}\bigg). \label{CSWFp}
	\end{eqnarray}
	Similarly, the LFWFs for $J^{z}=-\frac{1}{2}$ hadron in coordinate space can be expressed as
	\begin{eqnarray}
	\tilde{\psi}^{\downarrow X}_{+\frac{1}{2}}(x,\bfr)&=&i(r^{1}-i r^{2})  \frac{g M^{2}_{X}}{4\pi} (1-x)^{\frac{3}{2}} K_{0}\bigg(\bfr \sqrt{\mathcal{M}_{X}}\bigg), \nonumber \\
	\tilde{\psi}^{\downarrow X}_{-\frac{1}{2}}(x,\bfr)&=&(x \, M_{X}+m_{q}) \frac{\bfr}{\sqrt{\mathcal{M}_{X}}} \frac{g M^{2}_{X}}{4\pi} (1-x)^{\frac{3}{2}} K_{1}\bigg(\bfr \sqrt{\mathcal{M}_{X}}\bigg). \label{CSWFn}
	\end{eqnarray}
	Here, $K_{p}$ represents the Bessel function of the second kind of order $p$.

	With transverse coordinate $(x,\bfr)$, the charge distribution in the transverse coordinate space is given by
	\begin{eqnarray}
	P_{X}^{q}(x,\bfr)=\bigg[\tilde{\psi}^{\uparrow X \ast}_{+\frac{1}{2}}(x,\bfr) \,	\tilde{\psi}^{\uparrow X}_{+\frac{1}{2}}(x,\bfr)+\tilde{\psi}^{\uparrow X \ast}_{-\frac{1}{2}}(x,\bfr) \, \tilde{\psi}^{\uparrow X}_{-\frac{1}{2}}(x,\bfr)\bigg]. \label{ChDis}
	\end{eqnarray}
	On substituting the values of LFWFs from Eqs. (\ref{CSWFp}) and (\ref{CSWFn}) in Eq. (\ref{ChDis}), the explicit expression for  charge distribution comes out to be
	\begin{eqnarray}
	P_{X}^{q}(x,\bfr)=\frac{g^{2}M^{4}_{X}}{(4\pi)^2} (1-x)^{3} \bfr^{2} \bigg[\frac{(x \, M_{X}+m_{q})^2}{\mathcal{M}_{X}}K_{1}^{2}\bigg(\bfr \sqrt{\mathcal{M}_{X}}\bigg) + K_{0}^{2}\bigg(\bfr \sqrt{\mathcal{M}_{X}}\bigg) \bigg].
	\end{eqnarray}
    The distributions corresponding to quarks in coordinate space for proton are presented in Fig. \ref{fig13d3Cordp} and for different content of strangeness carrying baryons  in Figs. \ref{fig14d3Cords} and \ref{fig15d3Cordss}. On moving from $p$ to $\Sigma^{+}$ ($\Sigma^{o}$ or $\Lambda$), the amplitude for an active $u$ quark increases but a fall is observed while going from $\Sigma^{+}$ ($\Sigma^{o}$ or $\Lambda$) to $\Xi^{o}$. Similar fall is observed in amplitude for the active quark being a  $s$ quark however, the fall is less pronounced as compared to that when the active quark is a $u$ quark. The peaks are also shifted to lower values of longitudinal momentum fraction from $p$ to $\Sigma^{+}$ ($\Sigma^{o}$ or $\Lambda$), the amplitude for an active $u$ quark increases but a fall is observed while going from $\Sigma^{+}$ ($\Sigma^{o}$ or $\Lambda$) to $\Xi^{o}$. This clearly implies that the quarks in $\Xi^{o}$ always carry lesser fraction of longitudinal momentum than its spectator diquark which is heavy and carry more longitudinal momentum fraction. Comparing $u$ and $s$ either in $\Xi^{o}$ or in any other strange baryon reveals that $s$ quark carry more longitudinal momentum fraction than $u$ implying a larger contribution of massive $s$ quark with large momentum fraction in case of charge distribution.  
    
    In the direction of increasing $\bfr$, the fashion of lowering down of the charge distribution is observed for all baryons which are under consideration and this lowering down of charge distribution is very fast for $p$ having no strangeness whereas the charge distribution for baryons having strange particle(s) show a slow reduction with $\bfr$. It has been found that charge distribution decreases very slowly for  $\Sigma^{+}$ ($\Sigma^{o}$ or $\Lambda$) as compared to $\Xi^{o}$ which indicates that charge is more concentrated towards center in $\Xi^{o}$ after $p$. Among $\Sigma^{+}$, $\Sigma^{o}$ and $\Lambda$, it is clear from the distributions that with the increase in the mass of a baryon, the decrement of the charge distribution with increasing $\bfr$ becomes slower.  
    
    \begin{figure*}
    	\centering
    	\begin{minipage}[c]{0.98\textwidth}
    		\includegraphics[width=7.5cm]{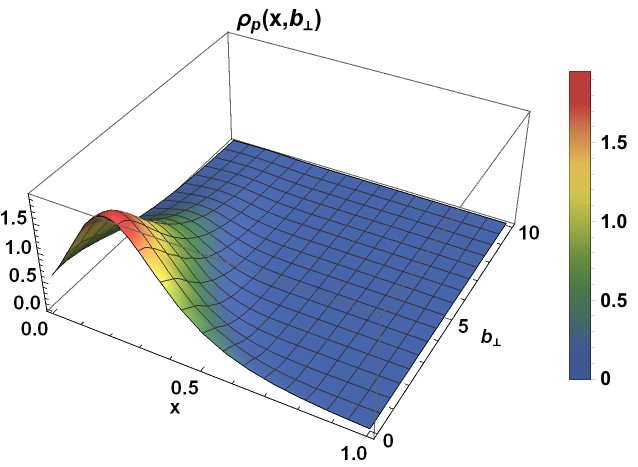}
    		\hspace{0.05cm}
    	\end{minipage}
    	\caption{\label{fig16d3Impp} (Color online) Charge  densities in impact parameter space for $u$ quark in proton.}
    \end{figure*}
     \begin{figure*}
    	\centering
    	\begin{minipage}[c]{0.98\textwidth}
    		(a)\includegraphics[width=7.5cm,clip]{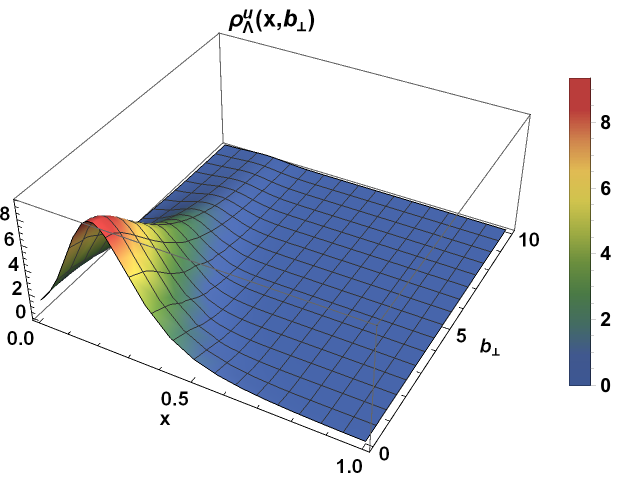}
    		\hspace{0.03cm}
    		(b)\includegraphics[width=7.5cm,clip]{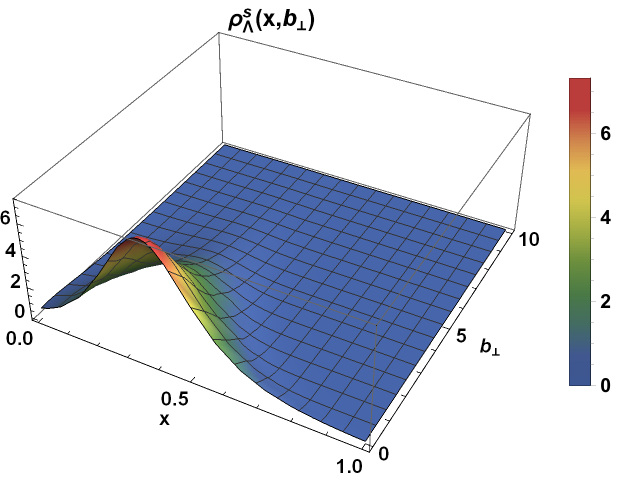}
    		\hspace{0.03cm}
    		(c)\includegraphics[width=7.5cm,clip]{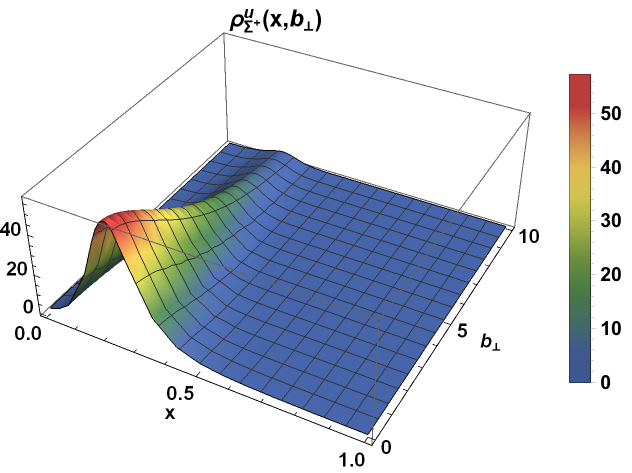}
    		\hspace{0.03cm}
    		(d)\includegraphics[width=7.5cm,clip]{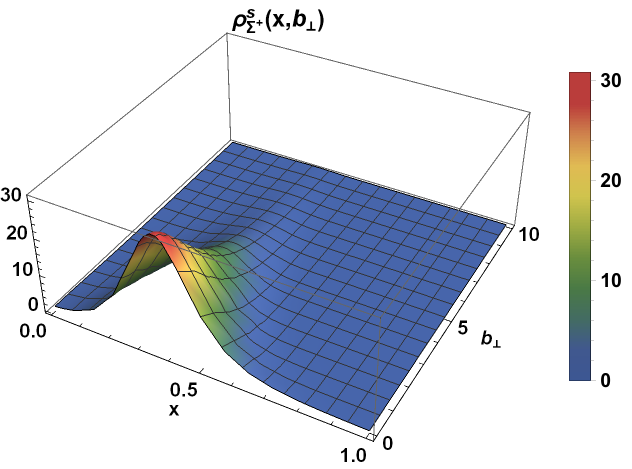}
    		\hspace{0.03cm}
    		(e)\includegraphics[width=7.5cm,clip]{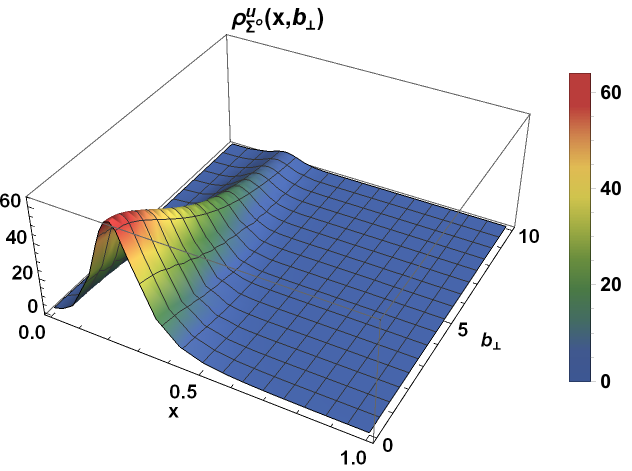}
    		\hspace{0.03cm}
    		(f)\includegraphics[width=7.5cm,clip]{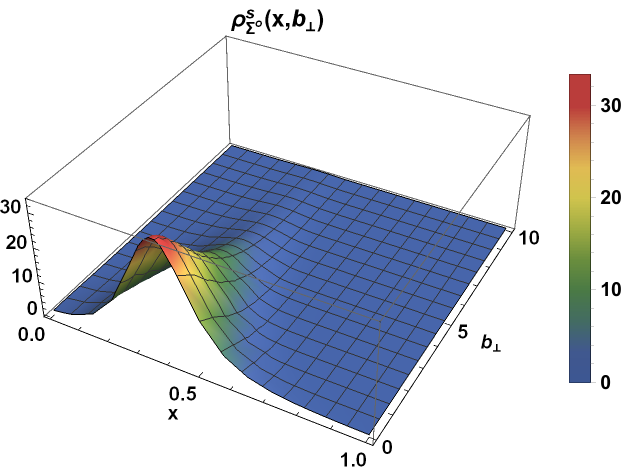}
    		\hspace{0.03cm} \\
    	\end{minipage}
    	\caption{\label{fig17d3Imps} (Color online) Charge densities in impact parameter space for baryons with single $s$ content. The left and right column correspond to $u$ and $s$ quarks sequentially.}
    \end{figure*}	
	Alternatively, Burkardt \cite{Burkardt:2002hr} showed that one can elicit  Dirac form factor $F_{1X}(\Dp)$ from the GPD $H_{X}^{q}(x,0,\Dp)$ and the Fourier transform of Dirac form factor interprets the charge distribution as a function of  transverse distance in the infinite momentum frame. Miller \cite{Miller:2007uy} assessed the parton charge densities through
	\begin{eqnarray}
	\rho_{X}^{q}(\bfb)=\int \frac{d^{2}\Dp}{(2\pi)^{2}} \, e^{-i \Dp \cdot \bfb} F_{1X}(\Dp).
	\end{eqnarray}
	Hence, in the impact parameter space, parton charge density in terms of GPD $H_{X}^{q}(x,0,\Dp)$ as a function of transverse distance and fraction of longitudinal momentum carried by an active quark \cite{Miller:2008jc} can be calculated using the relation
	\begin{eqnarray}
	\rho_{X}^{q}(x,b_{\perp})=\int \frac{d^{2}\Dp}{(2\pi)^2} \, e^{-i \Dp \cdot \bfb} H_{X}^{q}(x,0,\Dp) =\frac{1}{(1-x)^{2}} \, P_{X}^{q}\bigg(x,\frac{\bfb}{-1+x}\bigg).
	\label{ChImp}
	\end{eqnarray}
	By employing Eqs. (\ref{CSWFp})-(\ref{ChDis}), the explicit expression for charge density comes out to be
	\begin{eqnarray}
	\rho_{X}^{q}(x,\bfb)=\frac{g^{2}M^{4}_{X}}{(4\pi)^2} \frac{\bfb^{2}}{1-x} \bigg[\frac{(x \, M_{X}+m_{q})^2}{\mathcal{M}_{X}} \, K_{1}^{2}\bigg(\frac{\bfb}{-1+x} \sqrt{\mathcal{M}_{X}}\bigg) + K_{0}^{2}\bigg(\frac{\bfb}{-1+x} \sqrt{\mathcal{M}_{X}}\bigg) \bigg].
	\end{eqnarray}
   
	\begin{figure*}
		\centering
		\begin{minipage}[c]{0.98\textwidth}
		(a)\includegraphics[width=7.5cm,clip]{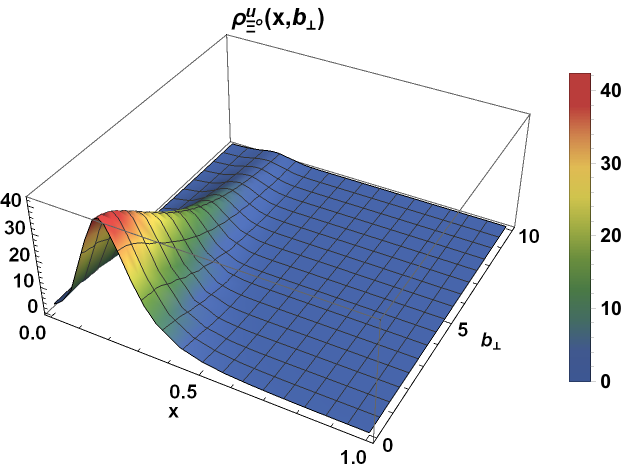}
		\hspace{0.03cm}
		(b)\includegraphics[width=7.5cm,clip]{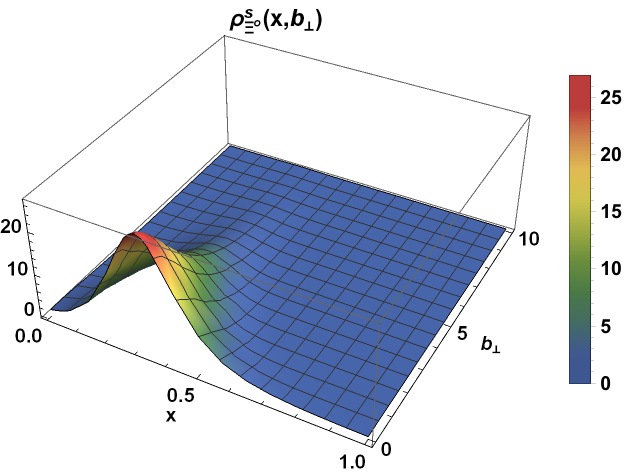}
		\hspace{0.03cm}
		\end{minipage}
		\caption{\label{fig18d3Impss} (Color online) Charge  densities in impact parameter space for baryon with double $s$ content. The left and right column correspond to $u$ and $s$ quarks sequentially.}
	\end{figure*}
	
	Charge Density in the impact parameter space for the case of proton is presented in Fig. \ref{fig16d3Impp} and corresponding to strange baryons, charge densities are presented in Figs. \ref{fig17d3Imps} and \ref{fig18d3Impss} respectively. While moving from $p$ to $\Sigma^{+}$ ($\Sigma^{o}$ or $\Lambda$), the amplitude of the distribution corresponding to an active $u$ quark increases with decrease in their widths which represents more constructive interference confined at a particular region of longitudinal momentum fraction for baryons carrying a single $s$ quark. On going to $\Xi^{o}$ from single $s$ carrying baryons, the amplitude of peak is found to decrease along with shifting of peak to lower fraction of longitudinal momentum. Similar trend is also observed for an active $s$ quark on moving from $p$ to $\Sigma^{+}$ ($\Sigma^{o}$ or $\Lambda$) and then to $\Xi^{o}$. The difference between an active $u$ and $s$ quark however, is a comparatively smaller amplitude of the distribution carried by the $s$ quarks with higher values of longitudinal fraction in the respective baryons. 
	
	\begin{figure*}
	\centering
	\begin{minipage}[c]{0.98\textwidth}
		\includegraphics[width=7.5cm]{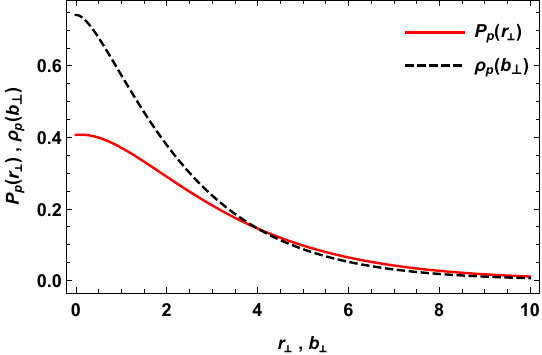}
		\hspace{0.05cm}
	\end{minipage}
	\caption{\label{fig19d3Compp} (Color online) Comparison of charge distribution in coordinate space and charge density in impact parameter space for $u$ quark in proton.}
\end{figure*}
\begin{figure*}
	\centering
	\begin{minipage}[c]{0.98\textwidth}
		(a)\includegraphics[width=7.5cm,clip]{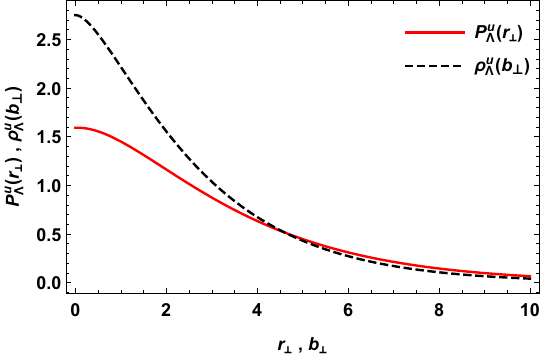}
		\hspace{0.03cm}
		(b)\includegraphics[width=7.5cm,clip]{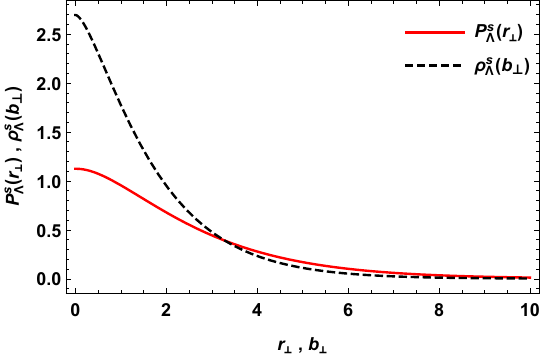}
		\hspace{0.03cm}
		(c)\includegraphics[width=7.5cm,clip]{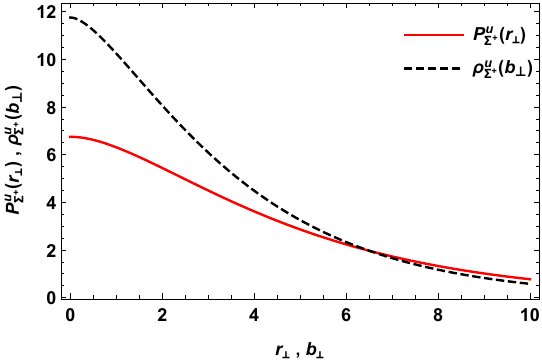}
		\hspace{0.03cm}
		(d)\includegraphics[width=7.5cm,clip]{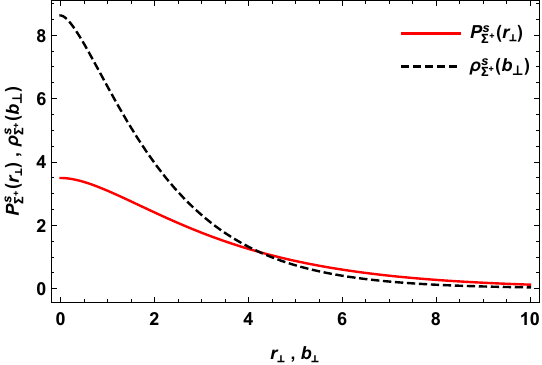}
		\hspace{0.03cm}
		(e)\includegraphics[width=7.5cm,clip]{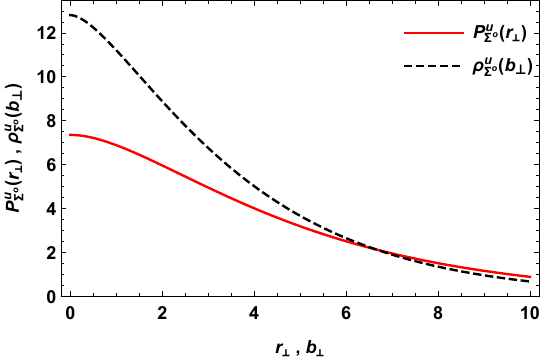}
		\hspace{0.03cm}
		(f)\includegraphics[width=7.5cm,clip]{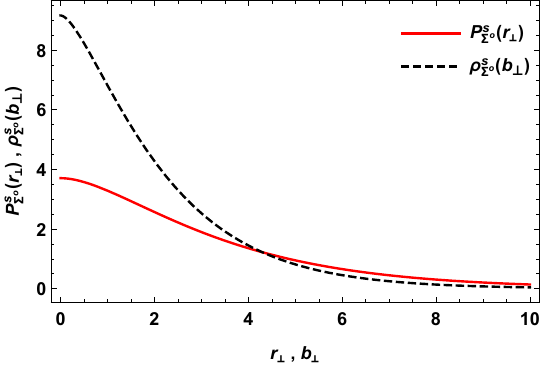}
		\hspace{0.03cm} \\
	\end{minipage}
	\caption{\label{fig20d3Comps} (Color online) Comparison of charge distributions in coordinate space and charge densities in impact parameter space for baryons with single $s$ content. The left and right column correspond to $u$ and $s$ quarks sequentially.}
\end{figure*}
\begin{figure*}
	\centering
	\begin{minipage}[c]{0.98\textwidth}
		(a)\includegraphics[width=7.5cm,clip]{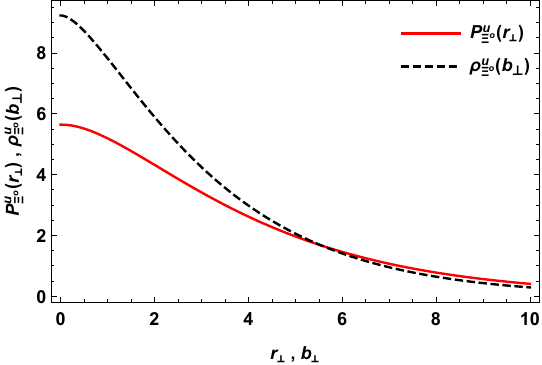}
		\hspace{0.03cm}
		(b)\includegraphics[width=7.5cm,clip]{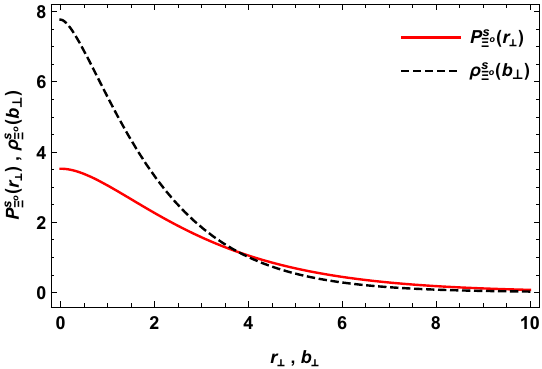}
		\hspace{0.03cm}\\
	\end{minipage}
	\caption{\label{fig21d3Compss} (Color online) Comparison of charge distributions in coordinate space and charge densities in impact Parameter Space for baryon with double $s$ content. The left and right column correspond to $u$ and $s$ quarks sequentially.}
\end{figure*}

	Variation with increasing $\bfb$ reveals that the charge density descends more rapidly for proton as compared to baryons having strangeness. In the queue, next particle is $\Xi^{o}$ whose charge density falls off smoothly with a shift of peak on smaller values of longitudinal momentum fraction. At last, we have $\Sigma^{+}$, $\Sigma^{o}$ and $\Lambda$ where the value falls down very slowly with $\bfb$. However, among these strange carrying baryons, the pace in decrement of distribution decreases with increase in the mass of a corresponding baryon. Further, comparing an active $u$ and $s$ quark, it has been found that $s$ quarks carry large longitudinal momentum fraction and fall off more frequently than $u$ in a respective baryon. 

	Moments in the momentum fraction $x$ lead to an informative description of charge densities in both the coordinate and impact parameter space. Two dimensional plots, representing a difference between charge distributions in coordinate space with coordinate $\bfr$ and in the impact parameter space with coordinate $\bfb$, have been presented in Figs. \ref{fig19d3Compp}, \ref{fig20d3Comps} and \ref{fig21d3Compss} for baryon having no strangeness, single and double $s$ content in baryons respectively. The vogue of decrease in charge distribution is almost the same in both the spaces and are indistinguishable in the region of high values of $\bfr$ and $\bfb$. The quantitative difference arises only at lower values of $\bfr$ and $\bfb$. The origin of this difference lies in the method to obtain the  expressions of charge densities in coordinate space and impact parameter space. In impact parameter space, charge density is derived from GPDs and we are dealing with skewedness GPDs $(\zeta=0)$. On fixing skewness parameter to zero, the longitudinal momentum fraction gets fixed and one can access the exact information about the charge distribution in impact parameter space. On the other hand, in the coordinate space, there is no way of fixing the skewness parameter or longitudinal momentum fraction as we derive the formula from an overlap form of wave functions well defined in coordinate space. As a consequence, the distribution may vary due to the involvement of longitudinal momentum. Another way of looking down at the slow decrease of densities in coordinate space is the presence of the factor $\frac{1}{(-1+x)}$ in Eq. (\ref{ChImp}). Charge density in coordinate space extends more towards outside than in impact parameter space as a consequence of this. 
	
	\section{Magnetization Density in Impact Parameter Space \label{secmagden}}
		\begin{figure*}
		\centering
		\begin{minipage}[c]{0.98\textwidth}
			\includegraphics[width=7.5cm]{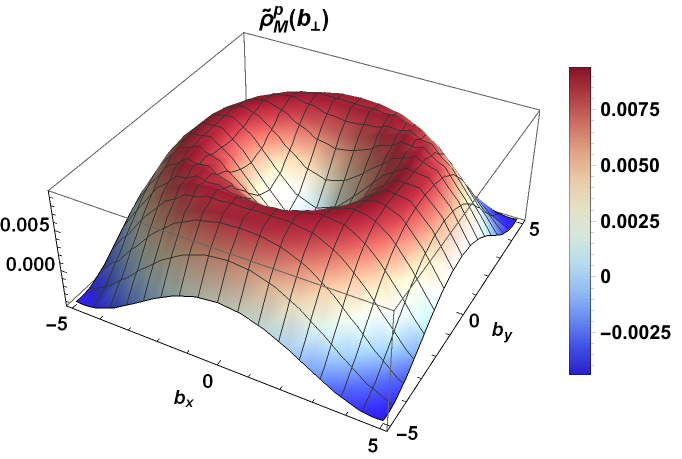}
			\hspace{0.05cm}
		\end{minipage}
		\caption{\label{fig22d3Magp} (Color online) Magnetization density in impact parameter space for $u$ quark in proton.}
	\end{figure*}
	Along the same lines of charge density in impact parameter space, one can elicit  Pauli form factor $F_{2X}(\Dp)$ from the GPD  $E_{X^{q}}(x,0,\Dp)$. This implies that the Fourier transform of the Pauli form factor will interpret the magnetic distribution as a function of  transverse distance in the infinite momentum frame \cite{Miller:2010nz}. We have
	\begin{eqnarray}
	\rho_{M}^{X_{q}}(\bfb)=\int \frac{d^{2}\Dp}{(2\pi)^{2}} e^{-i \Dp \cdot \bfb} F_{2X}(\Dp).
	\end{eqnarray}
	For anomalous magnetization density we have 
	\begin{eqnarray}
	\tilde \rho_{M}^{X_{q}}(\bfb)=-b_{y}\frac{\partial}{\partial b_{y}} \rho_{M}^{X_{q}}(\bfb).
	\end{eqnarray}
	By making use of the above equations, the anomalous magnetization density in terms of GPD $E_{X^{q}}(x,0,\Dp)$ can be calculated using
	\begin{eqnarray}
	\tilde \rho_{M}^{X_{q}}(\bfb)=\bfb Sin^{2}(\phi) \int \frac{\Dp^{2} d\Dp}{2\pi} J_{1}(\Dp \bfb) \int dx \, E_{X}^{q}(x,0,\Dp),
	\label{MagDen}
	\end{eqnarray}
	where $\phi$ is the angle between the direction of a hadron polarization (or the transverse magnetic field) and $\bfb$.
	
		\begin{figure*}
		\centering
		\begin{minipage}[c]{0.98\textwidth}
			(a)\includegraphics[width=7.5cm,clip]{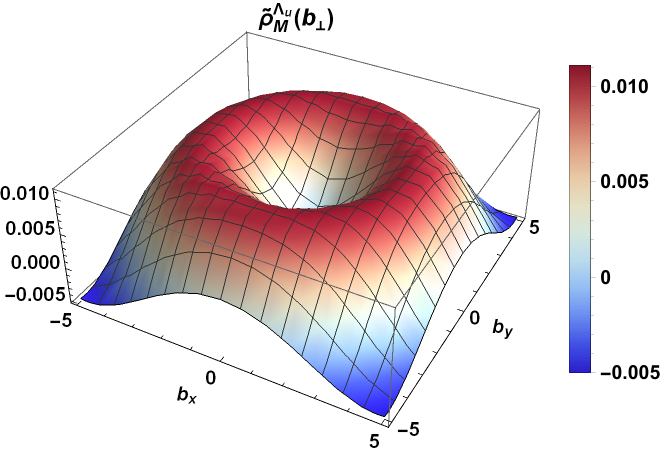}
			\hspace{0.03cm}
			(b)\includegraphics[width=7.5cm,clip]{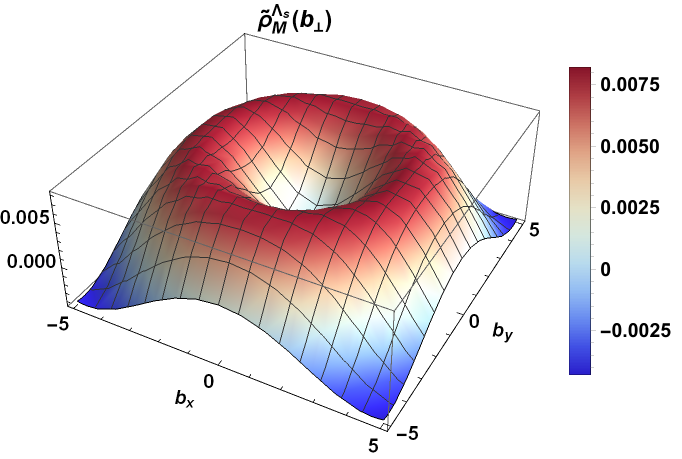}
			\hspace{0.03cm}
			(c)\includegraphics[width=7.5cm,clip]{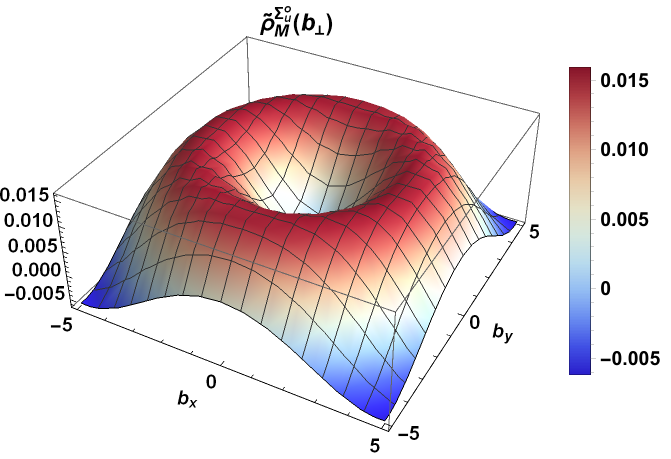}
			\hspace{0.03cm}
			(d)\includegraphics[width=7.5cm,clip]{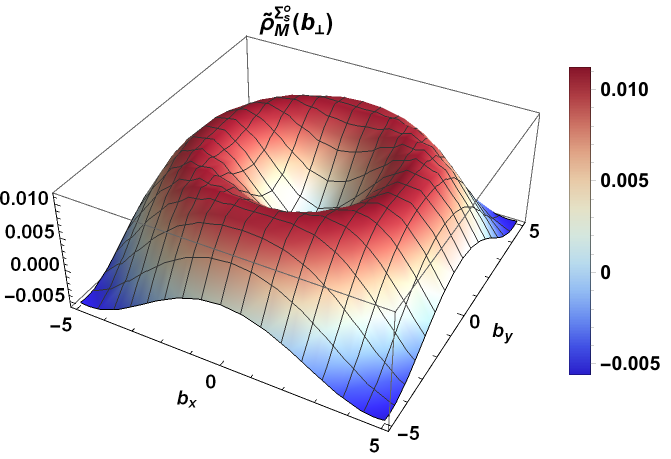}
			\hspace{0.03cm}
			(e)\includegraphics[width=7.5cm,clip]{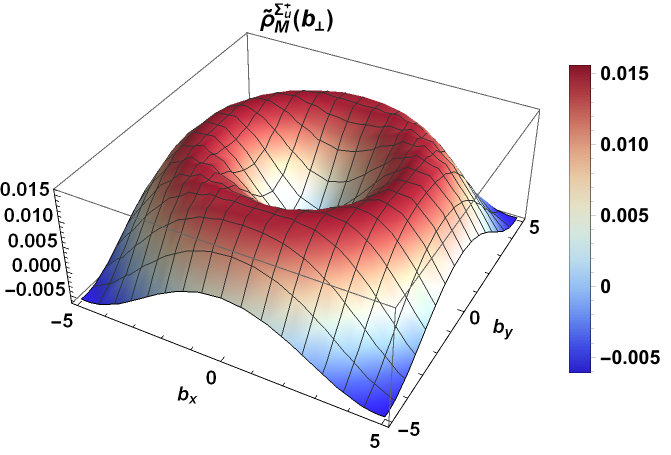}
			\hspace{0.03cm}
			(f)\includegraphics[width=7.5cm,clip]{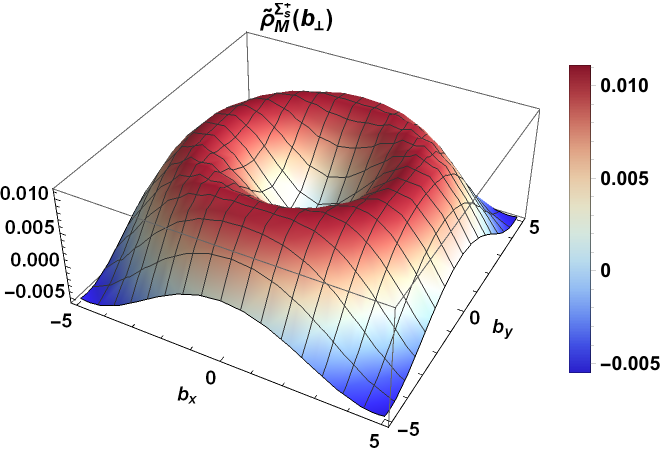}
			\hspace{0.03cm} \\
		\end{minipage}
		\caption{\label{fig23d3Mags} (Color online) Magnetization densities in impact parameter space for baryons with single $s$ content. The left and right column correspond to $u$ and $s$ quarks sequentially.}
	\end{figure*}
		\begin{figure*}
		\centering
		\begin{minipage}[c]{0.98\textwidth}
			(a)\includegraphics[width=7.5cm,clip]{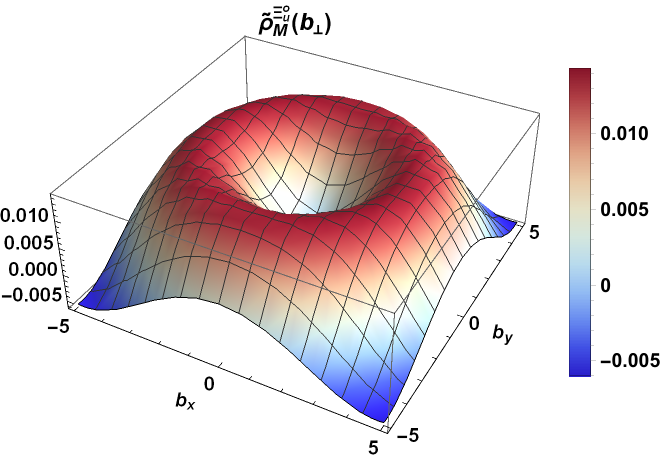}
			\hspace{0.03cm}
			(b)\includegraphics[width=7.5cm,clip]{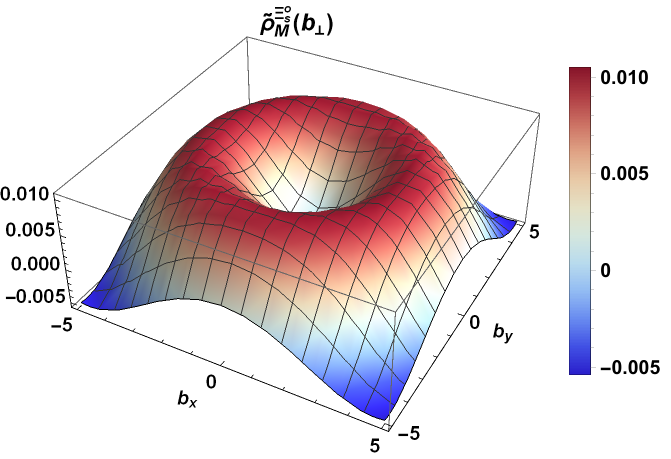}
			\hspace{0.03cm}\\
		\end{minipage}
		\caption{\label{fig24d3Magss} (Color online) Magnetization densities in impact parameter space for baryon with double $s$ content. The left and right column correspond to $u$ and $s$ quarks sequentially.}
	\end{figure*}
	The outcome of magnetization density calculated by using Eq. (\ref{MagDen}) for proton and strange baryons are portrayed in Figs. \ref{fig22d3Magp}, \ref{fig23d3Mags} and \ref{fig24d3Magss} respectively. It has been observed that the amplitude of the magnetization density increases as we jump from $p$ to $\Lambda$, $\Sigma^{o}$ and $\Sigma^{+}$. However, a decrease is observed for $\Xi^{o}$. The baryons $\Lambda$, $\Sigma^{o}$ and $\Sigma^{+}$ follow the fact that more the mass of a baryon, more is the amplitude of charge distribution. This variation is observed for both $u$ and $s$ quark. An expression of magnetization density has a dependency on an angle, $\phi$ and it gives zero anomalous magnetization density for $\phi=0$. In the present work, we have considered $\phi=\frac{\pi}{2}$ corresponding to which maximum value of an anomalous magnetic moment can be obtained.

	Values of maxima corresponding to the different baryons have been showed in Table \ref*{tab_peakv}. Second row of this table reveals that the magnetization density of an active $u$ quark among baryons carrying single $s$ quark increases with increase in the mass of a baryon whereas the magnetization density for baryon carrying two $s$ quarks decreases as compared to $\Sigma^{+}$ and $\Sigma^{o}$. On moving to third row, representing the maxima of magnetization density for an active $s$ quark, similar trend of increase and decrease in the values of $\bfb$ while going from $\Lambda$ to $\Xi^{o}$ is observed. The magnetization density is otherwise zero at $\bfb=0$, increases with increase in $\bfb$, reaches its maxima and then falls off. This trend is same for all particles irrespective of strangeness content. Comparing the distributions of an active $u$ and $s$ quark in any baryon, the maxima is found to lie at a smaller value of $\bfb$ for massive $s$ quark than light $u$ quark. This trend is same even in the baryon having single strange content.
	
		\begin{table}[h]
		\centering
		\begin{tabular}{|c|c|c|c|c|c|}
			\hline
			$\text{Particle (X)}  $~~&~~$ p  $~~&~~$  \Lambda $~~&~~$ \Sigma^{+} $~~&~~$ \Sigma^{o} $~~&~~$ \Xi^{o} $ \\
			\hline
			$|\bfb|_{u} $~~&~~$ 3.178 $~~&~~$ 3.185 $~~&~~$ 3.221 $~~&~~$ 3.224 $~~&~~$ 3.204 $ \\
			\hline
			$|\bfb|_{s} $~~&~~$ - $~~&~~$ 3.135 $~~&~~$ 3.155 $~~&~~$ 3.157 $~~&~~$ 3.148 $ \\
			\hline
		\end{tabular}
		\caption{Maxima of magnetization density for octet baryons.}
		\label{tab_peakv} 
	\end{table}
	
	\section{Conclusion}\label{seccon}
    To probe the internal structure of an identical spin-parity $J^{P}=(\frac{1}{2})^{+}$ members of an octet baryon, we used the quark-scalar diquark model to investigate the quark helicity independent GPDs. Depending on the different content of strangeness in a baryon, the behavior of different baryons such as $\Sigma^+$, $\Xi^o$, $\Lambda$ and $\Sigma^o$ was studied in the present work. 
	By employing a one-loop quantum fluctuations of the Yukawa theory, we formulated the explicit expressions for GPDs in the momentum as well as impact parameter space. The distributions presented in the momentum space for an unpolarized and transversely polarized quarks in an unpolarized state of members of baryon octet reveals that a massive strange quark always takes away more fraction of longitudinal momentum from its parent baryon as compared to a light quark. This fact is also concluded in context of the TMDs. This clearly implies that a massive quark always carries less kinetic energy than a light quark. On comparing the distributions of unpolarized and transversely polarized quarks, we have found that the transversely polarized quark always carries less fraction of longitudinal momentum than unpolarized quarks. 
	
	We have utilized the GPDs to study the probability densities of different distribution functions in impact parameter space to get a circumstantial visualization of the baryon states containing different content of strangeness. The 3D distributions in impact parameter space exhibiting the probability density corresponding to an active quark localized at a transverse position $\bfb$ with a longitudinal fraction momentum $x$ for the different members of a baryon octet reassure the results obtained in momentum space. Along with this, the probability densities are more intense when an active quark lies at the transverse center of a respective baryon with no extension in the longitudinal direction and a decrement in the probability densities has been observed with the shifting of distributions to smaller values of longitudinal momentum fraction carried an active quark. This is because the separation of an active quark from the center of momentum of a respective baryon increase in the transverse plane. 

	Further, transverse charge density representing the charge density of an active quark carrying a longitudinal momentum fraction $x$ at a particular transverse position $\bfb$ was studied. For all the octet baryons, it has been found that the profoundest amount of charge density is localized at the transverse center of a respective baryon and it decreases as the value of impact parameter increases. However, the pace of decrement of charge densities vary with the strange content in a baryon. For the case of proton, the charge density away from the center is the least. For $\Lambda$, $\Sigma^{o}$ and $\Sigma^{+}$, the charge distribution corresponding to an active quark in a transverse plane decreases with the increase in impact parameter at a slower rate in comparison to $\Xi^{o}$. To analyze the spatial structure of a baryon, we also presented the charge distributions in the coordinate space. The vogue of variation of charge distributions in coordinate space for baryons with different strangeness is same as that in impact parameter space with a difference in the presence of charge distribution even at larger transverse position from the center of momentum of a respective baryon. 
	
	GPD corresponding to transversely polarized quarks is employed to study the magnetization density in impact parameter space. From the quantitative as well as qualitative analysis of the distributions, magnetization density is found to be zero at $\bfb=0$. At a certain value of  $\bfb$, it gives a maxima which varies for different baryons depending on the strange content present in a respective baryon. An increase in the value of  $\bfb$ has been observed while moving from proton to baryon carrying single $s$ quark with  a decrease for the case of $\Xi^{o}$. In each baryon, we observed the  maxima lying at a smaller value of $\bfb$ for $s$ quark than $u$ quark. 
     
    To substantiate the results observed in our calculations, there is a need of a hard exclusive processes involving production of the strange baryons. By analyzing such processes, one can extract the CFFs by calculating the cross-section and asymmetries in the process as they are linear combinations of GPDs. In ALICE and STAR experiments, strange baryons are being investigated but since the life span of these strange baryons is small, cross-section is low and there are large backgrounds. With these challenges to access GPDs of strange baryons experimentally,  GPD measurements also demand collision energy significantly above the hadronic scale to allow transfer of energy and momentum \cite{Accardi:2023chb}. Since the meagerness of theoretical calculations and models introduce large uncertainties and vagueness in the extraction of experimental outcomes, this study will help to fill this gap.

    \section{Acknowledgement}
    H.D. would like to thank the Science and Engineering Research Board, Department of Science and Technology, Government of India through the grant (Ref No.TAR/2021/000157) under TARE scheme for financial support.

\end{document}